\documentclass[sigconf, sig-alternate-10pt]{acmart}

\usepackage{amsmath}
\usepackage{algorithm}
\usepackage{algorithmic}
\usepackage{subfigure}
\usepackage{xcolor}
\usepackage{pgfplots}
\usepackage[dvipsnames]{xcolor}
\pgfplotsset{compat=newest}

\usepackage{hyperref}
\usepackage{tikz}
\usepackage{subcaption}
\usepackage{enumitem}

\usepackage{multirow}

\newcommand{\specialcell}[2][c]{%
  \begin{tabular}[#1]{@{}c@{}}#2\end{tabular}}

\AtBeginDocument{%
  }

 \setcopyright{none}
\begin{document}
\pagestyle{empty}
\title{RACT: Retrieval Augmented Column-Table Learning and Prediction for Multi-Table Schema Matching}


\author{Leonard Traeger}
\affiliation{%
  \institution{University of Maryland,}
  \city{Baltimore County}
  \country{USA}}
\email{leonard.traeger@umbc.edu}

\author{Enas Khwaileh}
\affiliation{%
  \institution{Utrecht University}
  \city{Utrecht}
  \country{The Netherlands}}
\email{e.t.k.khwaileh@uu.nl}

\author{Andreas Behrend}
\affiliation{%
 \institution{Technical University of Cologne}
 \city{Cologne}
 \country{Germany}}
\email{andreas.behrend@th-koeln.de}

\author{George Karabatis}
\affiliation{%
  \institution{University of Maryland,}
  \city{Baltimore County}
  \country{USA}}
\email{georgek@umbc.edu}


\begin{abstract}
  Schema matching, a critical task for integrating data from diverse sources, seeks to identify correspondences between columns across different schemas. In multi-table (holistic) schema matching, columns with similar semantic meaning may reside in tables with different contexts due to heterogeneous schema designs, where similarity-based techniques are inadequate. The focus of this paper is exploiting referential context into schema matching by introducing \textit{RACT learning} and \textit{prediction}, a self-supervised framework enabling the \textit{probabilistic retrieval of candidate tables} for source columns to constrain relevant column candidates. Experiments demonstrate that this approach outperforms similarity-based baselines on matching multi-table schemas. In subsequent matching experiments, constraining the column search space via \textit{top-t} tables improves both average matching precision and completeness by up to $+70\%$. 
\end{abstract}


\maketitle




\section{Introduction}
\label{sec:intro}

Schema matching is a core process in data integration that aims to identify semantically related elements across heterogeneous data models, such as those found in enterprise data, cloud spaces, and marketplaces. The matching task alone is not trivial due to {\em linguistic} and \textit{design conflicts} between the schemas. Existing schema matching methods focus on two-table matching and encode column metadata (e.g., column name, data types, constraints) or the actual records (i.e., instance-based matching) into embeddings using pre-trained encoder-based language models. While these embeddings are highly effective at capturing direct semantic correspondences using similarities \cite{koutras_valentine_2021-1, hattasch_its_2022, tu_unicorn_2023, sheetrit_rematch_2024, zhang_smutf_2025, liu_magneto_2025}, they struggle to resolve \textit{contextual} matches, that is, matches across relational tables that convey different contexts.

\begin{figure}[t]
  \centering
  \includegraphics[width=.57\linewidth]{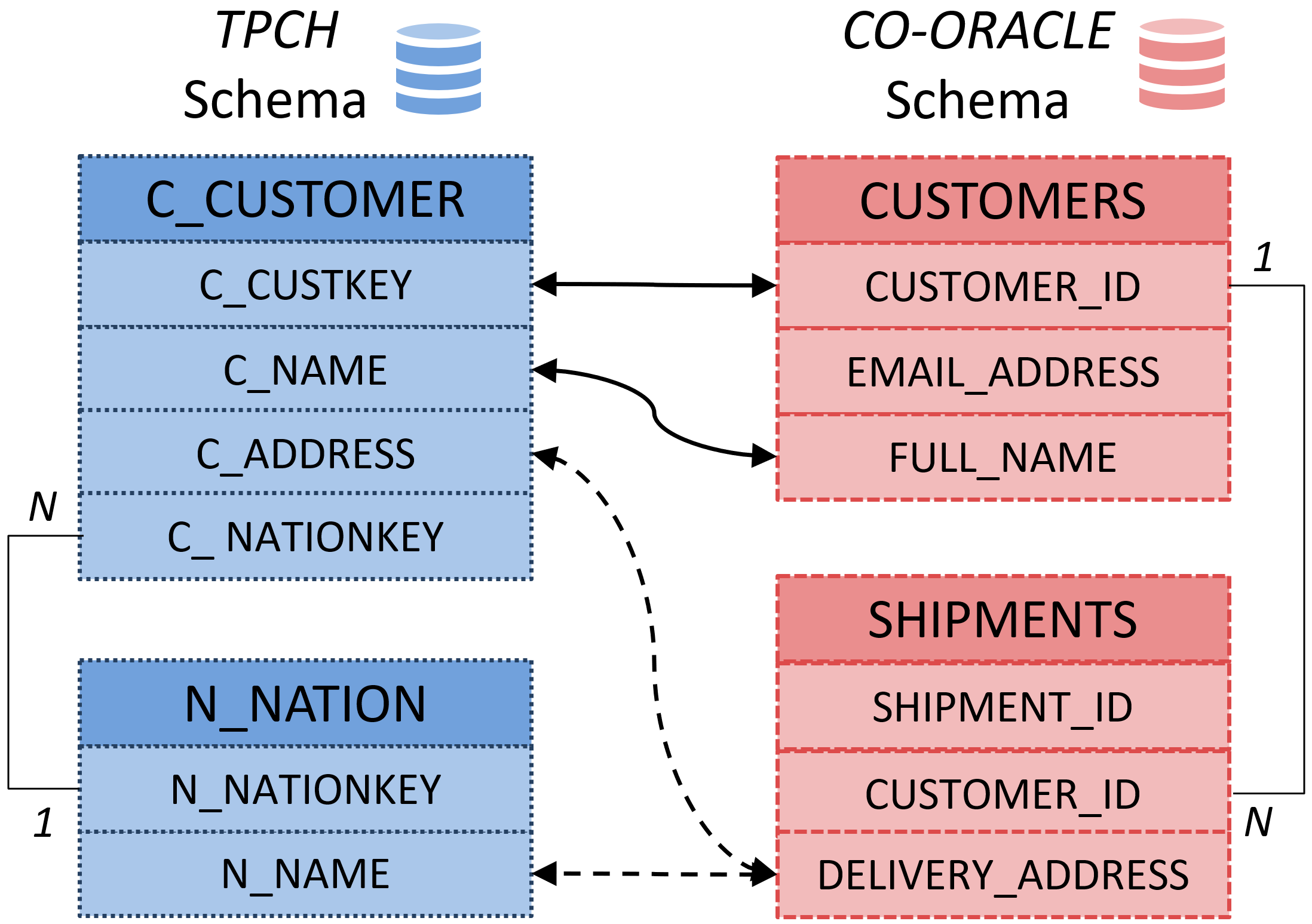}
  \caption{Example of direct (straight) and contextual (dashed) column matches in multi-table schema matching.}
  \label{fig:motivation}
\end{figure}

\textbf{Motivating example.} Relational schemas vary due to application-specific requirements that result in different logical design choices (e.g., ER-Modeling). For example, Figure \ref{fig:motivation} illustrates two schemas from the retail domain on customers and orders. The TPCH schema stores the customer entities along with addresses in its \verb|C_CUSTOMER| table with a separate country table \verb|N_NATION|. Conversely, schema CO-Oracle stores customer profiles in the table \verb|CUSTOMERS| and their logistics data separately in \verb|SHIPMENTS|.

Column alignments between the  tables \verb|C_CUSTOMER| and \verb|CUSTOMERS| are straightforward. For instance, \verb|C_CUSTKEY| and \verb|C_NAME| of table \verb|C_CUSTOMER| directly match (solid lines) to \verb|CUSTOMER_ID| and \verb|FULL_NAME| of table \verb|CUSTOMERS|. This occurs because the two tables are semantically equivalent, they share the same context, and for such cases,  comparing their entire metadata using similarities also yields high scores. Therefore, matching columns residing in tables with similar  contexts correctly encode into precise embeddings for similarity-based matching. 

On the contrary, column matches with different table contexts are more challenging. Consider, for example, the column pairs \verb|C_ADDRESS| from table \verb|C_CUSTOMER| and \verb|DELIVERY_ADDRESS| from table \verb|SHIPMENTS| (connected through a dashed line). Both represent an `address' related to a customer and they should match, even though their host tables differ in context. \verb|C_ADDRESS| is a property of the customer, whereas \verb|DELIVERY_ADDRESS| describes a logistical event, two completely different table contexts that both associate with a customer. Furthermore, considering that the \verb|DELIVERY_ADDRESS| column also contains country names, it should also be matched to TPCH's \verb|N_NAME| column in the countries table \verb|N_NATION| (second dashed line). 

\textbf{Problem:}  When we match columns that are contextually related to each other, they may sometimes belong to tables that are dissimilar. Consequently, their embeddings would be positioned away from each other, deeming them as non-matching. However, these contextually related columns should match despite their low embedding similarity. Unfortunately, all supervised \cite{ayala_leapme_2022, tu_unicorn_2023, zhang_smutf_2025}, active learning \cite{zhang_schema_2023, meduri_alfa_2024}, and LLM-based \cite{remadi_prompt_2024, sheetrit_rematch_2024, liu_magneto_2025} frameworks employ embedding similarity, resulting in erroneous matches. 

We need a matching solution that recognizes the context of each column based on existing referential constraints of its host table. Ideally, when matching TPCH's column \verb|C_ADDRESS| to the Oracle schema, one would first retrieve relevant tables such as \verb|CUSTOMERS| and its structurally bound \verb|SHIPMENTS| table and then better contextualize \verb|DELIVERY_ADDRESS| as a similar concept.

\textbf{In this paper, we propose a novel framework for matching relational schemas}: To overcome the problem of matching columns across tables with different contexts, we propose to first rank relevant candidate tables for source columns. Therefore, we transform each schema into a graph where the nodes represent tables and the edges represent referential constraints. Subsequently, we enrich columns with referential context from the graph and use their embeddings to self-supervise a neural model for host table prediction. Such learned models can be applied to other source schemas for constraining the search process from similarity (approximations) to target table predictions (probabilities). 

Figure \ref{fig:related_work_difference} illustrates our framework as a complementary preceding phase to classical matching. We probabilistically constrain the column vector space by ranking table candidates (dotted) while preserving classical column matching using similarities (straight). 

Our contributions are the following: 

\begin{figure}[t]
 \vspace{-3em}
  \centering  
  \includegraphics[width=0.47\textwidth]{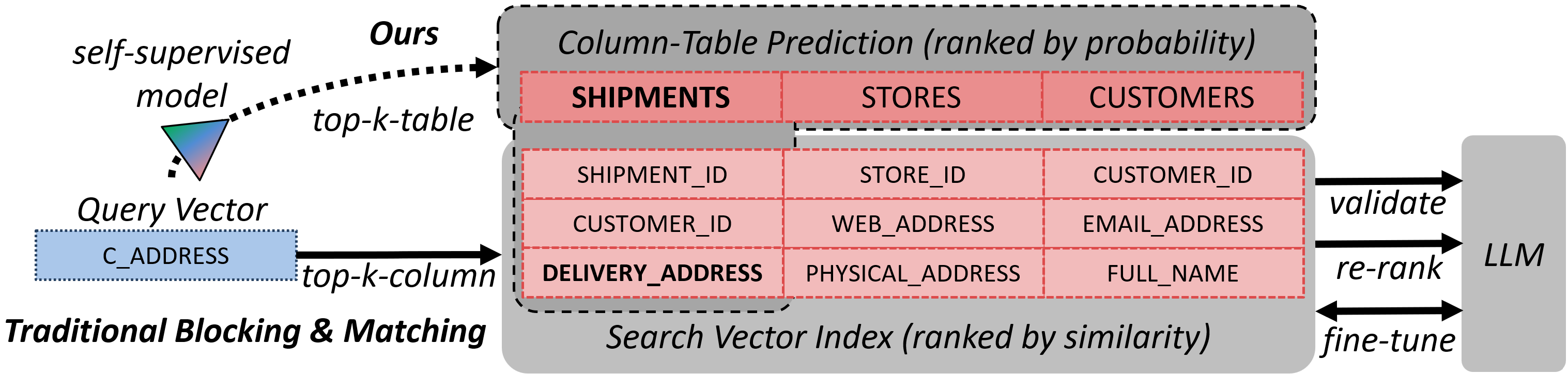}
  \caption{Schema Matching with Column-Table Prediction.}
  \label{fig:related_work_difference}
\end{figure}

\begin{itemize}
    \item A formal description of Column-Table Learning and Prediction to automatically navigate through relational schemas for holistic multi-table matching (Section \ref{sec:problem_formulation}).
    \item A technique to transform relational schemas into graphs for generating retrieval augmented columns (Section \ref{sec:method}.a).
    \item A novel self-supervised methodology to automatically predict \textit{top-t} table candidates, which contain the relevant target column for matching (Section \ref{sec:method}.b) 
    \item An empirical evaluation of table candidate retrieval to convey the practical effectiveness of the proposed framework (Section \ref{sec:evaluation}). We show that our self-supervised models excel at generic (up to $+13\%$) and larger (up to $+24\%$) target schemas compared to a similarity-based baseline. The Holistic approach remains robust as well as efficient because only one model is required for all schemas in a scenario.
    \item An ablation study of Column-Table Prediction for a Blocking and Matching pipeline that empirically validates an improvement in the completeness (recall) and quality (mAP) of schema matches (up to $+70\%$).
\end{itemize}

\begin{figure*}[t]
  \centering  
  \subfigure[\centering Retrieval Augmented Columns (Input Features and Learning Targets)]{\includegraphics[width=0.545\textwidth]{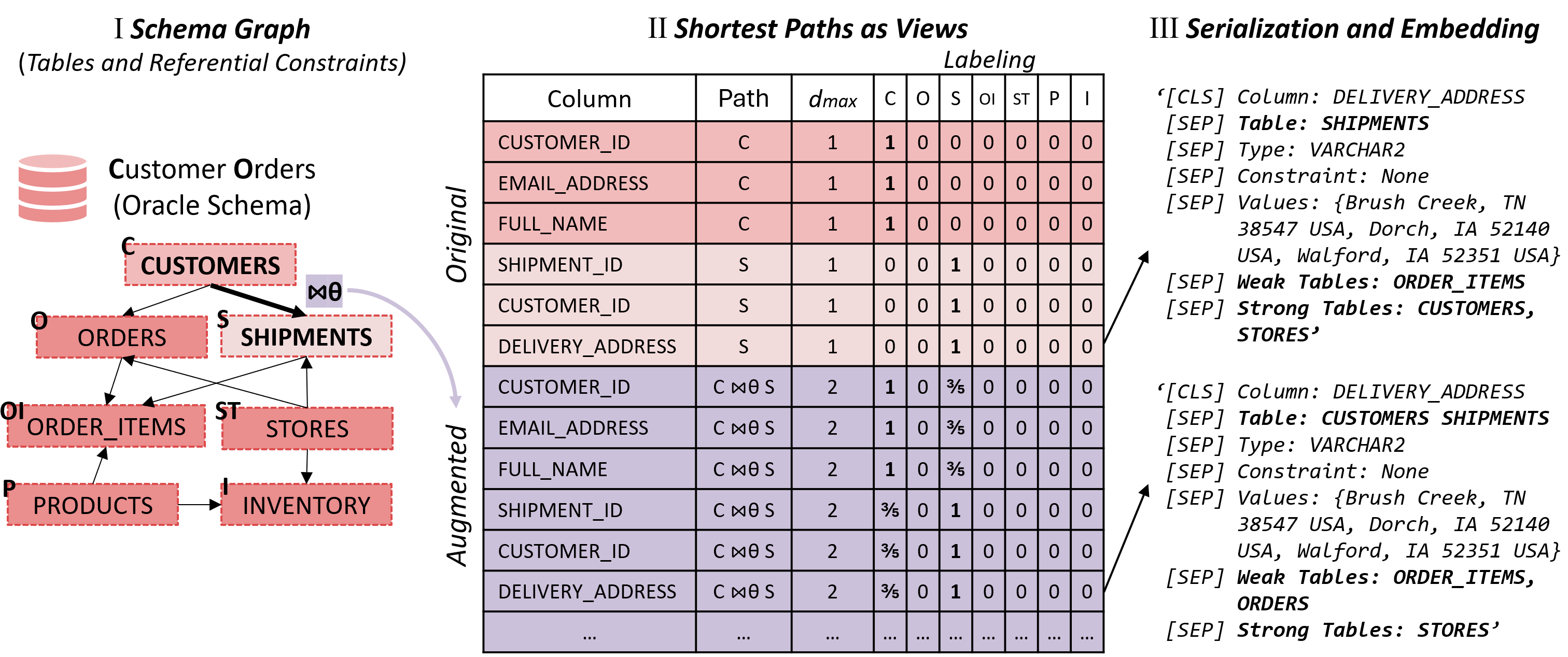}}
  \subfigure[\centering Self-Supervised Column-Table Learning (Training)   ]{\includegraphics[width=0.305\textwidth]{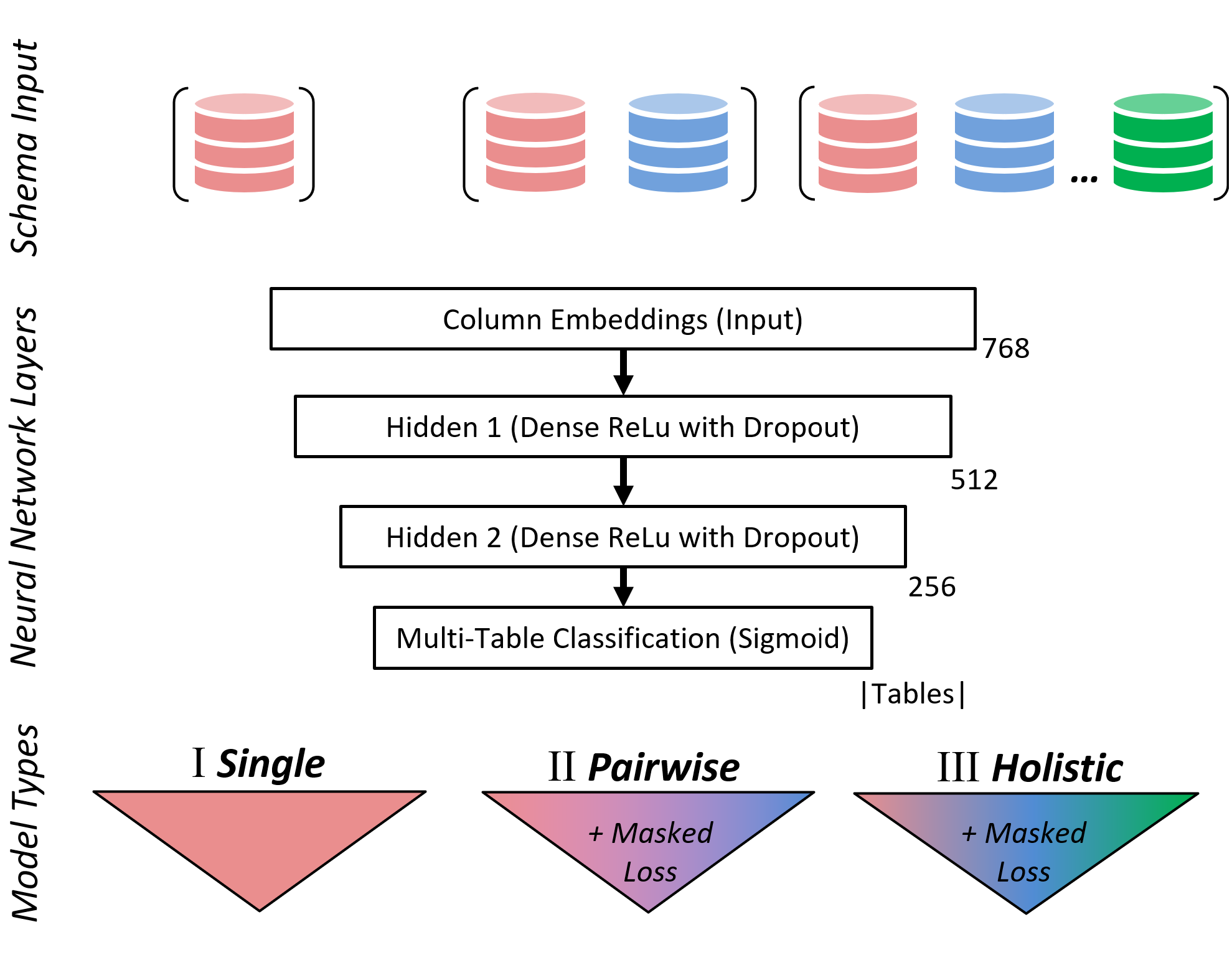}}
  \subfigure[\centering Column-Table \newline Prediction (Inference)]{\includegraphics[width=0.13\textwidth]{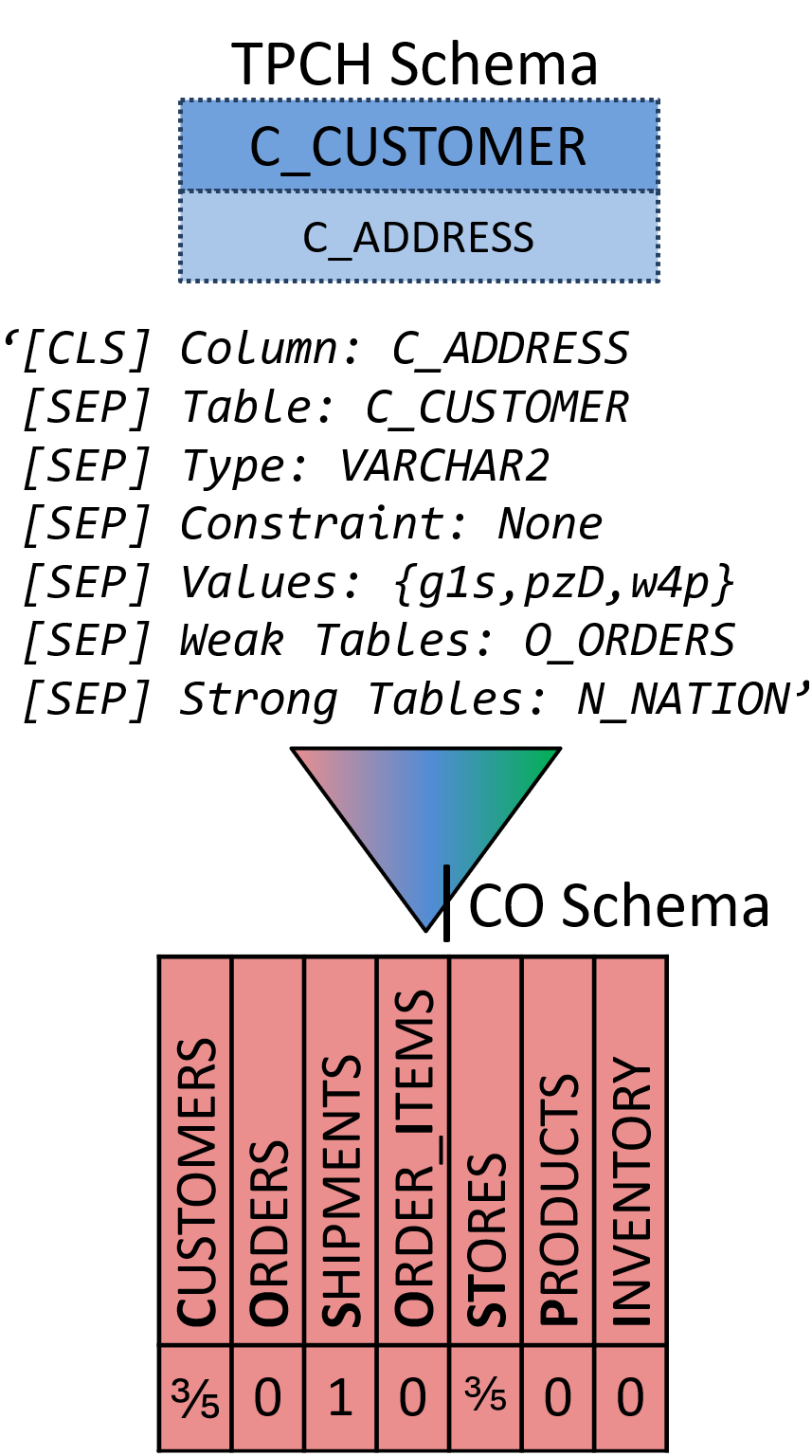}}
  \caption{RACT Learning and Prediction Framework for Schema Matching}
  \label{fig:method}
\end{figure*}

\section{Related Work}
\label{sec:related_work}

To the best of our knowledge, no prior work employs self-supervised learning for column-driven target table retrieval for holistic schema matching. Therefore, we review related work on traditional and language model-driven schema matching techniques. 

\noindent \textbf{Traditional Schema Matching.} Classical methods \cite{rahm_survey_2001} either use schema metadata linguistics like column names, data types, and constraints (schema-based) such as Cupid \cite{madhavan_generic_2001} or COMA \cite{do_coma_2002}. Notably, Similarity Flooding \cite{melnik_similarity_2002} converted relational schemas into graphs to propagate similarity. We also transform relational schemas into graphs, but for encoding referential context into column serializations. Alternatively, the overlap in column values (instance-based) signals similarity \cite{zhang_automatic_2011, cappuzzo_creating_2020, khatiwada_santos_2023}. While values provide meaningful column context useful for matching, value overlap primarily signals joinability \cite{zhu_josie_2019, subramaniam_numjoin_2023, koutras_omnimatch_2024}, which does not resolve contextual differences across tables.
We evaluate the impact of combining schema with instance-based matching using semantic embeddings in Section \ref{sec:evaluation}. The hybrid system COMA++ aggregates weighted similarities based on learning them from other validated scenarios \cite{aumueller_schema_2005}. However, we holistically self-supervise Column-Table patterns among multiple relational schemas in order to probabilistically constrain the search space before fine-grained matching. 

\noindent \textbf{Language Model-Driven Schema Matching.} In modern approaches \cite{castro_fernandez_seeping_2018, koutras_valentine_2021-1, suhara_annotating_2022, badaro_transformers_2023}, \textit{encoder-based language models} are trained \cite{cappuzzo_creating_2020} or employ pre-trained ones \cite{reimers_sentence-bert_2019} to transform serialized columns into embeddings. They are essential \cite{sheetrit_rematch_2024, liu_magneto_2025} for efficient and effective scaling down of multi-source matching tasks by quickly approximating exponential search spaces (via Blocking) for subsequent fine-grained ranking using similarities (via Matching with e.g., Cosine or LLM re-ranking). 


Hättasch et al. propose a two-step approach that first blocks candidate table pairs to then match column embeddings using similarity \cite{hattasch_its_2022}. Similarly, LLMATCH proposes a table selection phase that maps source-to-target tables \cite{wang_llmatch_2025}. Both techniques use separate table-table similarities to constrain the search space of column matching. In our approach, we bridge the conceptual column and table layers using Column-Table Learning.

Alternatively, LEAPME \cite{ayala_leapme_2022}, Unicorn \cite{tu_unicorn_2023}, and SMUTF \cite{zhang_smutf_2025} represent supervised techniques to calibrate matching weights from annotated scenarios. Furthermore, PoWareMatch \cite{shraga_powarematch_2021}, Zhang et al. \cite{zhang_schema_2023}, and Alfa \cite{meduri_alfa_2024} propose active learning frameworks to reduce human labeling cost. Remadi et al. propose to validate and rerank column matches using \textit{decoder-based language models} (LLM) \cite{remadi_prompt_2024}. Finally, Magneto extends LLMs as rerankers for fine-tuning encoder-based language models \cite{liu_magneto_2025}. \textit{Notably, each of these approaches adopts similarity-based techniques to derive candidate matches for its specialized matching framework. Our approach self-supervises Column-Table patterns as a preceding phase for complementing candidate retrieval.}

In this context, Sheetrit et al. recently proposed to approximate the search space from column-column to Column-Table retrieval using embedding similarity (ReMatch's \textit{Candidate Target Tables Retrieval}) \cite{sheetrit_rematch_2024}. Inspired by self-supervised learning \cite{traeger_collaborative_2026}, we propose self-supervising the Column-Table patterns of schemas for retrieving candidates for external source columns. This puts our probabilistic Column-Table framework in direct comparison with ReMatch's similarity-based retrieval, evaluated in Section \ref{sec:evaluation}.

\section{Problem Formulation}
\label{sec:problem_formulation}

We are given a set of relational schemas $S = \{S_{1}, S_{2}, \ldots, S_{k}\}$ that we refer to as a schema matching scenario. Each schema $S_{k} = \{t_{k_1}, t_{k_2},$ $\ldots, t_{k_i}\}$ contains a set of tables and each table $t_{k_{i}} = \{c_{k_1}, c_{k_2},$ $ \ldots, c_{k_j}\}$ contains a set of columns. For multi-source matching, the columns of one schema $S_k$ are aligned with at least another schema $S_m$. The alignment between the schemas is not fully bijective, since they may include one-to-one and one-to-many linkages as well as non-correspondences. 

\textit{Schema Matches}: We define the alignment between schemas as the set of column pairs between them: $M(S) = \{(c_{k_j}, c_{m_n}), \ldots: t_{k_i} \in S_k \land t_{m_l} \in S_m \land c_{k_j} \in t_{k_i} \land c_{m_n} \in t_{m_l} \}$ where $S_k, S_m \in S$ and $k \neq m$. A column pair $(c_{k_j} \cong c_{m_n})$ represents semantic congruence as a symmetric relationship. Note that the set of all truly matched (correctly aligned) column pairs  $M(S)$ is unavailable in practice but constitutes the ideal output set (ground truth) of matches.

\textbf{Definition 1.} \textit{Match Types:} To assess the capability of handling schemas with heterogeneous data models, we partition the matches $M(S)$ into two disjoint types based on the table contexts of columns:

\begin{enumerate}
    \item \textit{Direct Matches} ($M_{dir}$): Column matches $c_{k_j} \cong c_{m_n}$ (e.g., \verb|FULL_NAME| $\cong$ \verb|C_NAME|) where congruence is evident by the semantics of host tables $t_{k_i}$ and $t_{m_l}$ (e.g., \verb|CUSTOMERS| $\cong$ \verb|C_CUSTOMER|). Similarity-based methods typically succeed here. 
    \item \textit{Contextual Matches} ($M_{ctx}$): Column matches under differing normalization levels (e.g., vertical partitioning), where the host tables of the column match $c_{k_j} \cong c_{m_n}$ (e.g., \verb|DELIVERY_| \verb|ADDRESS| $\cong$ \verb|C_ADDRESS|) are not direct semantic counterparts (e.g., \verb|SHIPMENTS| $\neq$ \verb|C_CUSTOMER|). Instead, the host table $t_{k_i}$ of column $c_{k_j}$ has an adjacent table $t_{k_y}$, reachable via relationships (e.g., referential constraints), that represents the direct counterpart of the column $c_{m_n}$ pairs' table $t_{m_l}$ (e.g., \verb|SHIPMENTS| $\Join$ \verb|CUSTOMERS| $\cong$ \verb|C_CUSTOMER|). These kinds of matches require resolving the schema context (e.g., logical modeling via referential constraints).
\end{enumerate}

Intuitively, a domain expert does not search for matches by sequentially comparing a query column $c_{m_n}$ against every potential target column in the tables of schema $c_{k_j} \in t_{k_i} \in S_k$. Instead, they would follow a top-down hierarchical search by navigating through the logical schema $S_k$ (e.g., looking at the ER diagram). Then, they would identify relevant tables to filter the search space to a contextualized subset of candidate columns (ref. \cite{sheetrit_rematch_2024}). 

Inspired by this human-oriented search process, our goal is to replicate this semantic intuition automatically. In order to computationally navigate through the logical schema, we aim to learn the latent distribution between the columns (attributes) and how they belong to the tables (entity types).

\textbf{Definition 2.} \textit{Column-Table Learning.} Given a schema $S_k$, we learn a logical model $\mathcal{L}: c_{k_j} \to t_{k_i}$ that maps columns to their host tables. We formally define this as a function $f_{k}$ that predicts the probability distribution over the set of base tables $S_k = \{t_{k_1}, t_{k_2}, \dots, t_{k_i}\}$ given a column's semantic context:
\begin{equation*}
\label{eq:schema_learning}
f_k(c_{k_j}) = P(t_{k_i} \in S_k \mid c_{k_j})
\end{equation*}
where a high probability indicates that the column $c_{k_j}$ semantically belongs to table $t_{k_i}$.

The goal of our work is to learn such a function for each schema. With a learned Column-Table model, we aim to search for correspondences of columns in external schemas $c_{m_n} \in S_m$ by first predicting their \textit{top-t} tables before fine-grained \textit{top-k} column matching occurs. Consequently, we first predict a column's most contextually relevant tables and reduce the search space by eliminating false positive candidate columns. 


\section{Method}
\label{sec:method}

RACT Learning is a self-supervised framework that precedes classical schema matching by constraining candidate host tables for a column from an external schema. We illustrate our framework in Figure \ref{fig:method} that consists of the three sequential phases, i.e., \textit{(a) Retrieval Augmented Columns}, \textit{(b) Self-Supervised Column-Table Learning}, and \textit{(c) Column-Table Prediction} for more effective schema matching. In contrast to established matching pipelines, RACT encodes referential information into column embeddings as well as in the self-supervised learning task for predicting candidate tables.

\textbf{Overview.} In the first phase (a), we represent schemas as directed graphs for our retrieval augmentation procedure that encodes column context (input features) and table origins (learning targets). In the second phase (b), each schema self-supervises a feedforward neural network architecture to learn the latent distribution between columns to tables (training). In the last phase (c), we apply the learned models for semantic Column-Table Prediction (inference) among the schemas in order to constrain contextually relevant tables before fine-grained column matching.   

\textbf{(a) Retrieval Augmented Columns.} We propose an augmentation strategy inspired by \textit{Retrieval-Augmented Generation (RAG)} \cite{li_towards_2025}. However, instead of retrieving unstructured text to better answer a query, we generate relational paths between tables as views that we use for augmented column representations. For example, Figure \ref{fig:method}.a.I shows the serialized information of the original \verb|DELIVERY_ADDRESS| column in the Oracle schema. It does not contain any customer context which is a problem for matching. However, serializing the column from the generated view between \verb|CUSTOMER|$\Join$\verb|SHIPMENTS| adds customer context (table names concatenation $\rightarrow$ view name). Hence, our rationale for column augmentation is twofold: First, integration usually necessitates conjunctive queries (e.g., contextual addresses in Figure \ref{fig:motivation}). Consequently, join transformations in one schema $S_k$ (ref. `parallel schema realities' \cite{herrmann_living_2017}) may contribute to matching it to another schema $S_m$. Second, as we aim to implement Column-Table Learning (ref. Definition 2) with data-hungry neural networks in a robust way, we advocate for increasing the column training examples.

\underline{Algorithm Overview}: The complete procedure for generating referential-aware column embeddings (input features) and soft multi-table labels (target features) is outlined in Algorithm \ref{alg:view_augmentation}. 

\underline{(a.I) Relational Schema Graph (Lines 1-2)}: We intend to incorporate logical modeling concepts into the schema matching process. In practice, the primary (PK) and foreign key (FK) definitions can be easily extracted from the DDL or metadata (e.g., \verb|USER_CONSTRAINTS| in Oracle or \verb|INFORMATION_SCHEMA| in MySQL). Alternatively, they can be profiled via functional dependencies with methods \cite{abedjan_dfd_2014, papenbrock_functional_2015, koutras_omnimatch_2024, koehler_orthogonal_2025, maynou_freyja_2026} that are complementary but out of scope in this work. 

To generate referential-aware column embeddings, we start by representing a schema as a graph. Formally, we adopt the definition by Paganelli et al. on \textit{Schema Graphs} (introduced for computing Full Disjunctions) \cite{paganelli_parallelizing_2019}. However, we extend it with explicit graph directionality 
because our augmentation relies on join path traversals between tables. Specifically, we observe that undirected traversals reflect random (indiscriminate) join paths, which ignore the conceptual (Entity-Relationship) data model of a schema \cite{chen_entity-relationship_1976}. 

For example, consider the \verb|CO_ORACLE| schema where the transactional table \verb|SHIPMENTS| references both \verb|CUSTOMERS| and \verb|STORES|. An undirected traversal could generate a path \verb|CUSTOMERS| $\to$ \verb|SHIPMENTS| $\to$ \verb|STORES|, semantically clustering two independent entity types. By enforcing directionality of the ER hierarchy between strong and weak entity types, we prevent overlays between semantic concepts. The above traversal terminates at \verb|CUSTOMERS| $\to$ \verb|SHIPMENTS| so that generated views retain their semantic core concepts. Also, the number of generated views is reduced  at the same time.

\textbf{Definition 3.} \textit{Directional Schema Graph:} Given a schema with tables $S_{k} = \{t_{k_1}, t_{k_2}, \ldots, t_{k_i}\}$ and referential constraints $R_{k} = \{(t_\text{strong}$, $c_\text{strong}$, $t_\text{weak}$, $c_\text{weak})\}$ ($t_{\text{strong}} \neq t_{\text{weak}}$ self-references excluded), we create a directional graph $G_k$. For directionality, we strictly define $t_{\text{strong}}$ as the referenced table (contains primary key $c_{\text{strong}}$) and $t_{\text{weak}}$ as the referencing table (contains foreign key $c_{\text{weak}}$).

Based on these primitives, the \textit{Directional Schema Graph} $G_k = (V_k, E_k)$ is a graph where vertices $V_k$ represent the tables in $S_k$. The edges $E_k$ are derived from $R_k$ and the directionality parameter $dir$ where \textit{`directed'} reflects $1\to N$ expansion ($t_\text{strong} \to t_\text{weak}$), \textit{`inverse'} reflects $N \to 1$ expansion ($t_\text{weak} \to t_\text{strong}$ ), and \textit{`undirected'} represents simple connectivity (ref. \cite{paganelli_parallelizing_2019}).

\underline{(a.II) Shortest Paths as Views (Line 3)}: With the graph $G_k$ established (Line 1-2 in Algorithm \ref{alg:view_augmentation}), we identify join paths to generate views. Specifically, we apply shortest path traversal between $d_{max}$ maximum visited tables to control view explosion and maintain semantic core table concepts. Formally, we define a \textit{View Path} $p$ as an acyclic sequence of vertices (tables) traversing through edges (join paths) in $G_k$. Each path $p \in P_k$ (Line 3) represents a view with a custom join operator. We use $\Join\theta$ to retain all columns. 

\begin{algorithm}[t]
\caption{Retrieval Augmented Columns $\oplus$}
\begin{algorithmic}[1]
\REQUIRE $S_{k} = \{t_{k_1}, t_{k_2}, \ldots, t_{k_i}\}$ schema tables, $E$ language encoder,\\ $R_{k} = \{(t_\text{strong}$, $c_\text{strong}$, $t_\text{weak}$,$c_\text{weak})\}$ schema references, $dir = \{\text{directed}, \text{inverse}, \text{undirected}\}$ directionality, $d_\text{max}$ visited tables as view, $\lambda$ decay factor for multi-table labels
\ENSURE Retrieval Augmented Columns
\STATE $G_k \leftarrow \textit{InitGraphVertices(}\{t_{k_1}, t_{k_2}, \ldots, t_{k_i}\}\text{)}$
\STATE $G_k \leftarrow \textit{InitGraphEdges(}R_{k}, dir\text{)}$ \texttt{//Unweighted.}
\STATE $P_k = \{p_1, p_2, \ldots, p_z\} \leftarrow \textit{AllShortestPaths(}G_k, d_\text{max})$
\\\texttt{//Generate acyclic shortest paths between tables $p_z =\{t_{k_a}, \ldots, t_{k_b}\}$ as uniquely augmented views where $|p_z| \in [1:d_\text{max}]$. Includes singletons for original schema tables where $|p_z|=1$.}\\
\STATE $RAC_k \leftarrow \emptyset$
\FORALL{$p_z \in P_k$} 
\FORALL{$c_{k_j} \in t_{k_i} \in p_z$} 
\STATE $c_{k_i}^{p_z} \leftarrow \textit{Copy(}c_{k_i})$ \texttt{//Init augmented column.}\\
\STATE $c_{k_i}^{p_z}\text{.table\_name} \leftarrow \textit{ConcatNames(}p_z)$ 
\STATE $c_{k_i}^{p_z}\text{.weak\_tables}  \leftarrow \textit{GetAdjacentTables(}p_z, \text{weak}) \setminus p_z$
\STATE $c_{k_i}^{p_z}\text{.strong\_tables} \leftarrow \textit{GetAdjacentTables(}p_z, \text{strong}) \setminus p_z$
\STATE $\vec{x}_{k_i}^{p_z} \leftarrow E(c_{k_i}^{p_z})$ \texttt{//Embedding (input features).}\\
\STATE $\vec{y}_{k_i}^{p_z} \leftarrow \textit{SetMultiTableLabels(}c_{k_i}^{p_z}, \lambda)$ \texttt{//Exponential decay for table labeling (learning targets).}\\
\STATE $RAC_k \leftarrow RAC_k \cup \{(\vec{x}_{k_i}^{p_z}, \vec{y}_{k_i}^{p_z})\}$
\ENDFOR
\ENDFOR
\RETURN $RAC_k = (\vec{X}_k, \vec{Y}_k)$  
\end{algorithmic}
\label{alg:view_augmentation}
\end{algorithm}

\underline{(a.III) Serialization (Lines 7-10) and Embedding (Line 11):} 
First, we describe the process of transforming our retrieval augmented columns into input features used for Column-Table Learning (b). For every column in an augmented view, we retrieve the referential context according to its table composition and neighboring tables for its serialization. Then, we use pre-trained language encoders such as Sentence-BERT \cite{reimers_sentence-bert_2019} to transform the sequences of words (serialization) into an embedding, a fixed-sized latent vector.

Formally, given an encoder-based language model $E$ and some serialization $s = \{w_1, w_2, \ldots, w_d\}$, first each word $w_d \in s$ is encoded into a set of embeddings. Consolidated as a matrix, the encoder transforms it via pooling in order to output a single condensed embedding $\vec{e} \in \mathbb{R}^d$ that captures the serialization's meaning. 

Following the work in \textit{Unicorn} \cite{tu_unicorn_2023} and \textit{Magneto} \cite{liu_magneto_2025}, we serialize the columns in a similar way. In extension to their two-table matching objective, we additionally include the table name (i.e., \textit{ConcatNames}) into each column's serialization to contextualize column boundaries for learning multi-table associations. 
Based on the findings in \textit{Magneto}'s verbose serialization variant, we also add prefixes \textit{``Column:''}, \textit{``Table:''}, \textit{``Type:''}, \textit{``Constraint:''}, and \textit{``Values:''} which act as semantic anchors within the latent space of our neural network architecture. Similarly, we use the \textit{``[CLS]''} token as a serialization convention that indicates the start of a column and \textit{``[SEP]''} to control Sentence-BERT's pooling mechanism such that column metadata components are treated as discrete features. We consider the following serialization approaches:

\begin{equation*}
    \small{
    \begin{aligned}
    Ser_{\textit{schema}}(c_{k_j}^{p_z}) = & \textit{[CLS] Column: $c_{k_j}^{p_z}$.name [SEP] Table: $c_{k_j}^{p_z}$.table\_name}\\
      & \textit{[SEP] Type: $c_{k_j}^{p_z}$.type [SEP] Constraint: $c_{k_j}^{p_z}$.constraint}
  \end{aligned}
  }
\end{equation*}

$Ser_{\textit{schema}}$ represents the classical schema-based matching approach based on Rahm and Bernstein survey \cite{rahm_survey_2001}. Here, the\\$c_{k_j}^{p_z}\text{.constraint}$ explicitly labels key columns as \textit{``PRIMARY KEY''}, \textit{``FOREIGN KEY REFERENCES. $t_{\text{strong}}(c_{\text{strong}})$'}, or combined. Note that this referential context is only embedded for columns that represent keys. For example, only the column \verb|CUSTOMER_ID| in Oracle tables \verb|SHIPMENTS| contains \textit{``...[SEP] Constraint:} \verb|FOREIGN KEY| \verb|REFERENCES| \verb|CUSTOMERS| \verb|(CUSTOMER_ID)|\textit{''}. On the other hand, the column \verb|DELIVERY| \verb|_ADDRESS| in the same \verb|SHIPMENTS| table has no explicit \verb|CUSTOMER| contextualization (i.e. \textit{``...[SEP] Constraint: None''}). 

\begin{equation*}
    \small{
    \begin{aligned}
    Ser_{\textit{+values}}(c_{k_j}^{p_z}) = & \textit{[CLS] Column: $c_{k_j}^{p_z}$.name [SEP] Table: $c_{k_j}^{p_z}$.table\_name}\\
      & \textit{[SEP] Type: $c_{k_j}^{p_z}$.type [SEP] Constraint: $c_{k_j}^{p_z}$.constraint} \\
      & \textit{[SEP] Values: $c_{k_j}^{p_z}$.value$_1, \ldots,$ $c_{k_j}^{p_z}$.value$_5$}
  \end{aligned}
  }
\end{equation*}

$Ser_{\textit{+values}}$ extends the former serialization with column values (instances) (ref. \cite{liu_magneto_2025}), representing hybrid-based matching. To measure the impact of table names, we set $Ser_{\textit{magneto}}$ as $Ser_{\textit{+values}}$ without the table component. For both, we prevent noise from lengthy text fields (e.g., BLOBs in a \verb|DESCRIPTION| column) to not surpass Sentence-BERT's token limit of 512 by only using up to five first values and truncate them to a maximum of 300 tokens (ref. \cite{liu_magneto_2025}). While values provide context, due to distribution shifts, the embeddings of some columns that actually match may become dissimilar. For example, the matching columns \verb|FULL_NAME| (CO-Oracle) and \verb|C_NAME| (TPCH) likely contain disjoint sets of customer names although they refer to the same semantic concept. 
In such cases, matching with the schema-based approach would be more effective. 

\begin{equation*}
    \small{
    \begin{aligned}
    Ser_{\textit{+reference}}(c_{k_j}^{p_z}) = & \textit{[CLS] Column: $c_{k_j}^{p_z}$.name [SEP] Table: $c_{k_j}^{p_z}$.table\_name}\\
      & \textit{[SEP] Type: $c_{k_j}^{p_z}$.type [SEP] Constraint: $c_{k_j}^{p_z}$.constraint} \\
      & \textit{[SEP] Values: $c_{k_j}^{p_z}$.value$_1, \ldots,$ $c_{k_j}^{p_z}$.value$_5$} \\
      & \textit{[SEP] Weak Tables: $c_{k_j}^{p_z}$.weak\_tables}\\
      & \textit{[SEP] Strong Tables: $c_{k_j}^{p_z}$.strong\_tables}\\
  \end{aligned}
  }
\end{equation*}

$Ser_{\textit{+reference}}$ is one of our contributions. It further extends the above serialization approaches by adding the referential context of a column within a table by using the directed schema graph. It is particularly important for resolving ambiguity in our Column-Table Learning approach (ref. Definition 2). For instance, a domain-expert might mistakenly match TPCH's \verb|C_ADDRESS| from table \verb|CUSTOMERS| with CO-Oracle's \verb|WEB_ADDRESS| or \verb|PHYSICAL_ADDRESS| columns in table \verb|STORES|, as both are addresses (ref. Figure \ref{fig:related_work_difference}). \textit{Particularly for matching relational schemas, we need to consider the referential context for effective table retrieval.} Therefore, the function \textit{GetAdjacentTables} introduced in Algorithm \ref{alg:view_augmentation} retrieves the weak and strong tables of a column's host table or tables (in case of augmented view). 

For example, Figure \ref{fig:method}.a.III shows different serializations of Oracle's \verb|DELIVERY_ADDRESS| column. Both retrieve the referential context of weak entity types \textit{GetAdjacentTables(}(\verb|SHIPMENTS|, \verb|CUSTOMER|), \textit{weak)} $ \rightarrow$ \verb|ORDER_ITEMS| and \verb|ORDERS| that are similar to TPCH's \verb|C_ADDRESS| (i.e., \textit{GetAdjacentTables(}\verb|C_CUSTOMER|, \textit{weak)} $ \rightarrow$ \verb|O_ORDERS|), aligning the attribute pairs closer in the semantic vector space.

Note that the addition of the referential context is strictly derived from the table information of a column (i.e. \textit{GetAdjacentTables(}$c_{k_{i}}^{p_z}$.table\_name)). This way, the context of the column embeddings are forced to adjust to the table surroundings within the schema graph. However, some column semantics may result in convoluted context. For example, all columns in CO-Oracle's \verb|SHIPMENTS| table (e.g., \verb|SHIPMENT_ID|) are also contextualized with \verb|CUSTOMERS| even though they represent different concepts. We mitigate such cases via graph directionality and $d_{\text{max}}$-limited views.

Generally, when a column does not have constraints, values, or strong and weak tables, we concatenate \textit{“None”} to the corresponding prefix value. 
In Section \ref{sec:evaluation}, we study the individual impacts of the serialization variants for Column-Table Prediction and Matching.

\underline{(a.III) Labeling (Line 12)}: Finally, we now describe the generation of the learning targets for the columns. Note that any augmented column from a view is treated as a materialized combination of base tables. Hence, we only encode host tables into a multi-class vector.

Naively, we could encode it as a one-hot vector. However, the binary notion would normalize the original Column-Table semantics. 
For example, columns from a view path would equally belong to all of its traversed tables. Therefore, we quantify the table distances from the originating host table of an column with soft-labels.

In line 12, the function \textit{SetMultiTableLabels} generates soft labels for each augmented column's multi-table vector $\vec{y}_{k_i}^{p_z} \in [0,1]^{|S_k|}$. Let $t_{origin}$ be the table where column $c_{k_i}^{p_z}$ resides in the original schema, and $t_{k_i} \in p_z$ be any traversed table in its augmented view path. We assign a column belonging to a schema's tables $S_k$ using the exponential decay function as follows:

\begin{equation*} 
\vec{y}_{k_i}^{p_z} = 
\begin{cases} 
e^{-\lambda \cdot d(t_{origin}, t)} & \text{if } t \in p_z  \\
0 & \text{if } t \in (S_k \setminus p_z) 
\end{cases} 
\end{equation*}

\noindent where $d(t_{origin}, t)$ returns the originating table distance in the schema graph $G_k$ and $\lambda \in (0, 1)$ is the decay hyperparameter.

The intuition behind soft labels is a means for balancing the independent column semantics while learning the referential schema context, controlled via $\lambda$ uniformly set to $0.5$. For example, the column \verb|DELIVERY_ADDRESS| from the augmented view path between the tables \verb|CUSTOMERS| (distance 1) and \verb|SHIPMENTS| (origin) results in the target vector $\{$\verb|CUSTOMERS|=$0.606$, \verb|STORE|=$0$, \verb|SHIPMENTS|=$1$, \dots\}. 

\begin{table*}
\caption{RACT Dataset with Retrieval Augmented Columns by Graph Directionality and Shortest Path Depth.}
\label{tab:scenarios_augmentation_statistics}
\begin{tabular}{ccccccccc}
\toprule
\textbf{Schema} & \specialcell{\textbf{\#Tab./\#Col.}} & \specialcell{\textbf{$dir$ectionality}}  & \specialcell{$d_\text{max}$=2}  & \specialcell{$d_\text{max}$=3} & \specialcell{$d_\text{max}$=4} & \specialcell{$d_\text{max}$=5} & \specialcell{$d_\text{max}$=6} & \specialcell{$d_\text{max}$=7} \\ 
\midrule
\multirow{2}{*}{CO-Oracle} & \multirow{2}{*}{7 / 43} & directed/inv. & 16 / 156 & 18 / 192 & - & - & - & - \\
 &  & undirected & 25 / 269 & 45 / 654 & 49 / 746 & - & - & - \\ \hline
\multirow{2}{*}{CM-MySQL} & \multirow{2}{*}{8 / 59} & directed/inv. & 15 / 173 & 20 / 299 & 23 / 403 & 24 / 445 & - & - \\
 &  & undirected & 22 / 287 & 36 / 629 & 48 / 1013 & 56 / 1333 & 62 / 1611 & 64 / 1721 \\ \hline
\multirow{2}{*}{TPCH} & \multirow{2}{*}{8 / 61} & directed/inv. & 16 / 180 & 23 / 337 & 26 / 417 & 27 / 457 & - & - \\ 
 &  & undirected & 24 / 299 & 44 / 753 & 58 / 1177 & 64 / 1393 & - & - \\ \hline
 \multirow{2}{*}{Sakila} & \multirow{2}{*}{15 / 70} & directed/inv. & 36 / 312 & 52 / 583 & 60 / 753 & 64 / 852 & - & - \\ 
 &  & undirected & 55	/ 530 & 101 / 1312 & 145 / 2249 & 187 / 3366 & 211 / 4104 & 221 / 4426 \\ 
 \bottomrule

 \multicolumn{8}{l}{The \textbf{\#Tab./\#Col.} column provides the original (i.e., $d_{\text{max}}$=1) number of tables and columns of a schema.}\\
 \multicolumn{8}{l}{Each $d_{\text{max}}$ column provides the number of views (shortest path traversals $|P_k|$) and augmented columns of a schema.}\\
\end{tabular} 
\end{table*}

\textbf{(b) Self-Supervised Column-Table Learning.} In this phase, we implement the learning objective established in Definition 2. We instantiate the function $f_k$ using a Multilayer Perceptron (MLP) in order to learn the non-linear latent space between columns and tables in a schema. Therefore, the model takes the retrieval augmented column embeddings and processes them through two fully connected hidden layers. The final layer applies a Sigmoid activation to predict the independent probability of a column belonging to each table in $S_k$. Before presenting our three model types \textit{(b.I) Single, (b.II) Pairwise,} and \textit{(b.III) Holistic}, we first elaborate on a general loss modification for our learning task.

\textit{Focal Loss for Recall}: Even with augmented path views, a column's multi-table-label vector will associate membership with a small fraction of tables (sparse positives), while the vast majority are irrelevant (dense negatives). Consequently, the learning targets $\vec{y}$ contain predominantly zeros. Standard loss functions (e.g., Binary Cross Entropy) will tend to converge to the trivial solution of predicting no table memberships at all in order to maximize accuracy. However, with our retrieval objective (c) that is similar to Blocking, we prefer a model that recommends plausible but incorrect tables over entirely missed ones. Therefore, we adopt \textit{Focal Loss} \cite{lin_focal_2020} to counter extreme class imbalances. By down-weighting the losses of `easy' negatives (e.g., \verb|CUSTOMER_ID| does not belong to \verb|PRODUCTS|), we force the gradient updates to focus on the sparse positive table labels (e.g., correctly map \verb|EMAIL_ADDRESS| to \verb|CUSTOMERS|).

\underline{(b.I) Single.} The naive approach is training a MLP model $f_k$ independently for each schema $S_k$ by using only its column embeddings $\vec{X}_k \in RAC_k$. However, the optimization of $f_k$ is non-trivial. Generally, the learned schema function $f_k$ should perform well on predicting its target tables $\vec{Y}_k \in RAC_k$ for its own columns (training phase). Secondly, the model should effectively predict candidate host tables for external $S_m$ schema columns (inference phase). 

Due to heterogeneity in nomenclature and normalization, the input column embeddings of one schema $S_k$ differ from those of another schema $S_m$. Therefore, a learned function $f_k$ based on (b.I) will generalize less effectively to external schemas unless the matching columns embed nearly identically. Referring to the example provided in Figure \ref{fig:motivation}, a model type (b.I) may still accurately predict correct host tables for semantically similar encoded columns. However, it will struggle predicting contextual ones as it cannot generalize to external columns that it has not seen during training.

\underline{(b.II) Pairwise.} To improve generalization, we propose training a MLP model on the \textit{union of input features} $\vec{X}_{km} \leftarrow \vec{X}_k \cup \vec{X}_m$ and the concatenation of learning targets $\vec{Y}_{km} \leftarrow (\vec{y}_k, \vec{y}_m) | \vec{y}_k \in \vec{Y}_k \land \vec{y}_m \in \vec{Y}_m$ from two schemas. Consequently, we expose the model to the column embeddings and multi-table labels of both schemas that we intend to match. Notably, we also increase the number of input columns (training examples) naturally. However, the learning targets (complexity) increase as well.

\textit{Learning with Masked Loss}: Since we only have access to the multi-table labels of columns from their originating schema, we employ masked loss over the undefined table labels from the other schemas. Preciously, when training one column embedding of one schema $\vec{X}_{k}$, we mask the prediction loss of the undefined table labels from the other schema $\vec{Y}_{m}$ and vice versa. Consequently, the model is not penalized for the unknown table labels while the shared layers still learn generalized features.

Note that the pairwise approach generalizes among columns between two schemas, but the number of model trainings increases quadratically with the number of schemas in a scenario $O(|S|^2)$.

\underline{(b.III) Holistic.} In the context of enterprises, cloud spaces, and data marketplaces, matching scenarios often involve more than two schemas $|S|>2$. To achieve scalability, we extend the pairwise approach (b.II) to a holistic one that unions the inputs and concatenates the targets for all schemas in the matching scenario $S = \{S_1, S_2, \ldots, S_k\}$. This enables the model to learn universal Column-Table patterns across all schemas, effectively reducing the number of required models in a scenario from $O(|S|^2)$ (b.II) to $O(1)$. However, this shared learned model type increases the training data and learning target dimensionality. Nonetheless, we consider its input scale and complexity favorable to neural learning.

\textbf{(c) Column-Table Prediction.} Once the Column-Table model $f$ is learned via either of the b.I, b.II, or b.III model types, we deploy it to predict the table context for external schema columns. 

Given a set of query columns from a source schema $S_k$, we first encode the original columns of base tables (classical matching) into embeddings $\vec{X}_k$ (ref. Phase (a) at $d_{max}$=1). We then perform a forward pass using the trained model $f$ to compute the probability distribution over the target tables in $S_m$. For each query column $c_{k_j}$, the model outputs a likelihood score for the base tables $t_{m_l} \in S_m$:
\begin{equation*} 
\hat{\mathbf{y}}_m = \sigma(f(\vec{x}_{k_j})) \big|_{S_m}
\end{equation*}
where $\hat{\mathbf{y}}_m\in [0,1]^{|S_m|}$ represents the vector of predicted probabilities of column $c_{k_j}$ belonging to the semantic context of the tables in $S_m$. Note that for pairwise (b.II) and holistic (b.III) model types, this vector is a projection (slice) of the table indices of schema $S_m$.

Instead of a strict threshold (e.g., $> 0.5$ for Focal Loss in (b)), we rank the target tables by their predicted probability scores. Then, we constrain the \textit{top-t} tables with the highest scores as the candidate set. Consequently, the traditional pipeline with column blocking (i.e., \textit{top-k} columns) and matching (e.g., cosine similarity) is restricted exclusively to the columns of the \textit{top-t} table candidates. 


Note that we aim to retrieve those tables needed to resolve both direct and contextual matches. While our referential-aware columns and pairwise/holistic model types aim to optimize high recall at lower \textit{top-t} table cardinalities for candidate retrieval, they  may still introduce false negatives and impact fine-grained matching.


\section{Evaluation}
\label{sec:evaluation}

In this section, we evaluate our RACT Column-Table Learning and Prediction framework as well as its impact on traditional schema matching pipelines. We first describe the experimental setup and introduce the evaluation metrics. 
Overall, we observe that: 
\begin{enumerate}
    \item Our self-supervised Pairwise and Holistic model types consistently achieve higher mean recall \textit{@top-t} tables than ReMatch's similarity-based baseline for both generic (up to $+13\%$) and larger (up to $+24\%$) target schemas.
    \item In an ablation study, \textit{@top-t} table prediction improves \textit{@top-k} column matching in both recall and mAP (up to $+70\%$).
\end{enumerate}

\subsection{Experimental Setup}

With our RACT framework, we focus on holistically matching \textit{relational schemas} with each other, each containing multiple tables that are connected via \textit{referential constraints}. This way, we are able to evaluate Algorithm \ref{alg:view_augmentation} and all proposed model types (ref. b.I-b.III). To this end, we curated a dataset consisting of four open-source relational schemas with semantically meaningful metadata from the retail domain:
CO-Oracle\footnote{Oracle sample schemas: \url{https://github.com/oracle-samples/db-sample-schemas}}, CM-MySQL\footnote{Classicmodels (MySQL tutorial schema): \url{https://www.mysqltutorial.org/getting-started-with-mysql/mysql-sample-database/}}, TPCH\footnote{TPCH \url{https://www.tpc.org/TPCH/} implemented at scale-factor=1 via DuckDB}, and Sakila\footnote{Sakila (MySQL sample schema): \url{https://dev.mysql.com/doc/Sakila/en/}}. 

Limitations of existing benchmarks: Unfortunately, common schema matching benchmarks such as those provided by the Valentine project \cite{koutras_valentine_2021-1} are not suitable for evaluating RACT because they consist of only table-to-table scenarios; matching a single source table to a single target table. This two-table setting applies to Wikidata\footnote{Wikidata Musicians: \url{https://www.wikidata.org/}}, ChEMBL, Magellan\footnote{Magellan: \url{https://sites.google.com/site/anhaidgroup/useful-stuff/}}, Magneto's biomedical GDC \cite{liu_magneto_2025}, and SMUTF's HDXSM \cite{zhang_smutf_2025}. Finally, we acknowledge recent multi-table matching scenarios in healthcare alignment by ReMatch \cite{sheetrit_rematch_2024} and LLMatch \cite{wang_llmatch_2025}. However, all these scenarios match source schemas to the single OMOP\footnote{OMOP Common Data Model: \url{https://www.ohdsi.org/data-standardization/}} model $(S_s \rightarrow S_t)$ and not holistically among all of them $(S_1 \leftrightarrow S_2 \leftrightarrow \ldots \leftrightarrow S_k)$ as we do.


\noindent \textbf{Retrieval Augmented Columns.}  Table \ref{tab:scenarios_augmentation_statistics} displays an overview of the number of tables and columns of each schema. Additionally, we show the number of views and augmented columns of each schema at maximum traversed tables $d_{\text{max}}$ value (ref. Algorithm \ref{alg:view_augmentation}). Note that the augmentation via \textit{directed} or \textit{inverse} graph directionalities is identical. Conversely, \textit{undirected} graph traversals generate nearly as twice more views and augmented columns.
For the CO-Oracle, CM-MySQL, and TPCH schemas, the augmentation of columns increases at a similar rate due to comparable table numbers (seven to eight). However, the  Sakila schema (15 tables) returns nearly twice as many view paths. In our evaluation, we employ RACT learning with original ($d_{\text{max}}$=1) and \textit{directed} $d_{\text{max}}$=2 augmentation. We exclude \textit{undirected} and depth $d_{\text{max}}\geq3$ parameter values as they mix semantic core table concepts and have shown to introduce noise to the learning task.

\begin{table}[b]
\centering
\caption{Scenarios and $M(S)$ Annotated Matches by Type.}
\label{tab:match_types_stats}
\begin{tabular}{lccc}
\toprule
Matching Scenario & Direct ($M_{dir}$) & Contextual ($M_{ctx}$) & $\sum$ \\
\midrule
{\small CO-Oracle   $\leftrightarrow$ CM-MySQL}& 17 {\small(63\%)}& 10 {\small(37\%)}& 27 \\
{\small CO-Oracle  $\leftrightarrow$ TPCH }& 23 {\small(92\%)} & 2 {\small(8\%)} & 25 \\
{\small CO-Oracle  $\leftrightarrow$  Sakila}& 15 {\small(54\%)} & 13 {\small(46\%)} & 28 \\
{\small CM-MySQL  $\leftrightarrow$ TPCH }& 24 {\small(80\%)} & 6 {\small(20\%)} & 30 \\
{\small CM-MySQL  $\leftrightarrow$ Sakila }& 22 {\small(56\%)} & 17 {\small(44\%)} & 39 \\
{\small TPCH $\leftrightarrow$ Sakila }& 22 {\small(65\%)} & 12 {\small(35\%)} & 34 \\
\bottomrule
\end{tabular}
\end{table}

\noindent \textbf{Annotated Matches by Type.} The ground truth was manually annotated by analyzing the ER diagrams and data samples, verified by two external data scientists. In Table \ref{tab:match_types_stats}, we provide an overview of the symmetric column matches between the six matching scenarios, categorized as direct or contextual matches. Note that each schema pair alignment presents its unique challenges. For instance, 
CO-Oracle (i.e., \verb|SHIPMENTS|) and Sakila (i.e., \verb|ADDRESS|) store addresses in separate tables from customers.

Incorporating referential context is needed to disambiguate column matches that yield high similarity scores. For example, the \verb|OFFICE| table in CM-MySQL and the \verb|SUPPLIER| table in TPCH share nearly identical columns (e.g., \verb|CODE|, \verb|ADDRESS|, \verb|PHONE|). However, their referential contexts reveal distinct concepts. \verb|OFFICE| is referenced by \verb|EMPLOYEES| and \verb|CUSTOMERS| (sales context), whereas \verb|SUPPLIER| is linked to \verb|PARTSUPP| and \verb|LINEITEM| (manufacturing context). Hence, we did not annotate any matches between them. 

Lastly, while Sakila also models the retail domain, it specializes on film rentals. While it contains straightforward matches for its columns in \verb|CUSTOMER|, \verb|ADDRESS|, and \verb|STORE| tables to the corresponding columns in the other retail schemas, we need to consider its referential semantics to correctly align the film rental tables with the other retail schemas.  For example, TPCH$\leftrightarrow$Sakila contains column matches between \verb|ORDER|$\simeq$\verb|RENTAL| and \verb|PART|$\simeq$\verb|FILM| tables.

\noindent \textbf{Serialization and Embedding.} We test all serialization variants in our RACT framework and encode them using Sentence-BERT\footnote{Sentence-BERT (all-mpnet-base-v2) is reported as the best general purpose model (\url{https://www.sbert.net/docs/sentence_transformer/pretrained_models.html}).} \cite{zhang_smutf_2025, liu_magneto_2025}. For RACT Learning and Prediction, we employ $Ser_{\textit{+reference}}$ serialization as it yielded the highest validation accuracy.

\noindent \textbf{Column-Table Learning Hyperparameters:} We train the Single (b.I), Pairwise (b.II), and Holistic (b.III) model types with homogeneous neural network architectures. The MLP network comprises a densely connected 768 $\to$ 512 $\to$ 256 architecture with dropout (0.2) and ReLU activations along with a final Sigmoid layer that predicts the multi-table vector. While these and other hyperparameters could be fine-tuned individually for each model type based on schema input and target complexity, we keep hyperparameters constant to ensure a controlled evaluation environment between the architectural strategies.

As we aim for a model that accurately captures all Column-Table patterns among the schemas (training), for each model type, we split the input column embeddings into 80\% (train) 20\% (validation) using stratified sampling and learn a 5-fold cross-validation ensemble. This way, we ensure generalization among the schemas across the entire column population at the cost of increased training time. 

We set the batch size to 16 to ensure granular gradient updates from individual column embeddings. We set both the Adam learning rate and L2-regularization to a low 0.001 value in order to ensure the model has the flexibility to learn subtle distinctions. Generally, we aim for a model that retains column-identity specificity (corresponds to classical similarity matching) while projecting it into the latent table space. Stronger regularization suppresses subtle cues that are needed to distinguish nearly identical \verb|CUSTOMER_ID| embeddings in the table  \verb|CUSTOMERS| from one in \verb|SHIPMENTS|.

Finally, we perform a grid-search for the $\alpha$ parameter in our adapted Focal Loss. We uniformly set $\alpha$=0.9, as higher values consistently improved validation accuracy by focusing on the sparse positive classes (correct table or table composition of view paths). The higher weight to the positive class aligns with our recall-oriented blocking objective for Column-Table Prediction.

\noindent \textbf{Column-Table Baseline.} We implement ReMatch's successful similarity-based \textit{Target Table Retrieval}  \cite{sheetrit_rematch_2024}. Their approach is computationally efficient as it does not require additional model training but heavily relies on data catalogs with rich descriptions\footnote{MIMIC-OMOP healthcare matches: \url{https://github.com/meniData1/MIMIC_2_OMOP}}. While LLMs could synthesize column (e.g., SEALM \cite{traeger_sealm_2025} or SMUTF \cite{zhang_smutf_2025}) or table descriptions, accurately capturing contextual matches (e.g., TPCH's \verb|PART_KEY| and \verb|ORDERDATE| to Sakila's \verb|FILM_ID| and \verb|RENTAL_DATE|) inherently requires the referential schema context that our framework provides. We leave LLM-integration for future work and focus in this evaluation strictly on the retrieval mechanism itself, excluding orthogonal data augmentation. We replicate ReMatch by using $Ser_{\textit{schema}}(c_{k_j})$ for source columns and adapt their table document structure into the following serialization: 

\begin{equation*}
    \small{
    \begin{aligned}
    Ser_{\textit{ReMatch}}^t(t_{m_l}) = & \textit{[CLS] Table: $t_{m_l}$.name}\\
      & \textit{[SEP] Primary Keys: \{$c$\textit{.name} $c$\textit{.type} $\mid c \in t_{m_l}$.\textit{pks}}\}\\
      & \textit{[SEP] Foreign Keys: \{$c$\textit{.name} $c$\textit{.type} $c$\textit{.ref} $\mid c \in t_{m_l}$.\textit{fks}}\}\\
      & \textit{[SEP] Columns: \{$c$\textit{.name} $c$\textit{.type} $\mid c \in t_{m_l}$.\textit{other}}\}
  \end{aligned}
  }
\end{equation*}

\noindent \textbf{Blocking and Matching.} For Blocking, we implement embedding-based retrieval using the FAISS library \cite{johnson_billion-scale_2021}. For each scenario, we build an IndexFlatL2 (Exact Nearest Neighbor) for interchanging target schema to efficiently retrieve similar \textit{top-k}=\{1, 2, 3, 5, 10, 20\} column candidates of the source schema. Note that while our ground truth matches $M(\{S_1, \ldots, S_k\})$ represent symmetric relationships between the schemas, the candidate sets retrieved via blocking differ based on the source (query items) and target (search items) schema assignments. For Matching, we compute the Cosine similarity for the candidates, ranked in descending order.

\subsection{Evaluation Approach and Metrics} 
\label{sec:evaluation_metrics}

We evaluate our RACT framework in two stages. First, we measure the performance of our RACT Prediction models (b.I-III) against ReMatch \cite{sheetrit_rematch_2024}. Second, we measure its impact on similarity-based schema Blocking and Matching as an ablation study.

\noindent \textbf{Column-Table Prediction.} We evaluate it as a semantic blocking mechanism. Therefore, we measure whether a source column successfully retrieves the host table of the matching target column. To this end, we adapt the annotated ground truth of schema matches $M(S)$ (ref. Section \ref{sec:problem_formulation}) to Column-Table pairs $M'(S)$. For every column pair $(c_{k_j}, c_{m_n}) \in M(S)$, we derive the target table $t_{m_l}$ where the matching column $c_{m_n}$ resides:

$M'(S) = \{(c_{k_j}, t_{m_l}) | \exists c_{m_n} : (c_{k_j}, c_{m_n}) \in M(S) \land c_{m_n} \in t_{m_l}\}$ 

We report \textit{Recall@top-t}, defined as the proportion of query columns $c_{k_j}$ for which the host table $t_{m_l}$ appears in the \textit{top-t} table predictions sorted by the model's output probabilities $\hat{\mathbf{y}}_m$. 

\noindent \textbf{Ablation Study for Schema Matching.} Secondly, we evaluate the impact of constraining the search space via \textit{@top-t} table prediction on classical schema matching pipelines. Note that the maximum \textit{@t} table cardinality value is equivalent to searching through the columns from all tables (no RACT Prediction $\equiv$ baseline). 

First, we compute the standard \textit{Recall@top-k} column metric \cite{bellahsene_evaluating_2011, koutras_valentine_2021-1, liu_magneto_2025}. Secondly, we compute the Mean-Average Precision (mAP) \textit{mAP@top-k} column metric that extends classical precision. Specifically, it considers whether all of the true linkages tend to get ranked highly as an overall score for recommending relevant matches. Note that we compute mAP exclusively for query columns that have a match in the ground truth $M(S)$, thereby isolating the ranking performance for linkable columns.

\subsection{Results}

\textbf{Column-Table Prediction.} The \textit{Recall@top-t} scores of each pairwise matching scenario are reported in Table \ref{tab:recall_performance} using  $Ser_{\textit{+reference}}$ serialization. To ensure validity, we trained each model type over ten independent runs and provide the mean recall performance with standard deviation at each \textit{top-t} table cardinality. 
For each scenario, we compare our \textit{Pairwise} and \textit{Holistic} model types against the two baselines: the naive \textit{Single} model and the similarity-based \textit{ReMatch} retrieval.

Without column augmentation, the Single model type contains insufficient training examples, which is why we only report the mean recall performance with retrieval augmented columns at $d_{\text{max}}$=2. However, for the Pairwise and Holistic model types, we naturally increased the training samples by merging the columns between two or more schemas. 
To discuss the model type performances among all scenarios, we cluster them into three categories based on the table prediction complexity:

\underline{Mean(Sakila$\neq$target)}: First, we discuss the mean recall of all scenarios with CO-Oracle, CM-MySQL, or the TPCH as target schemas due to the similar number of target tables (seven to eight). At \textit{@top-t}=1, Single(2) reaches 0.454 and ReMatch 0.573 recall on average, whereas both the Holistic(2) (0.628) and Pairwise(1) (0.647) models outperform the naive Single ($+43\%$) and ReMatch baseline ($+13\%$). At \textit{@top-t}=2, the Pairwise(2) model with augmented columns slightly overtakes the un-augmented one from 0.771 to 0.785, both on par with ReMatch (0.787). The Holistic(2) model exceeds both baselines by $+3\%$ (0.81). 

Notably, we observe that ReMatch performs competitively in scenarios that predominantly contain direct matches (ref. Table \ref{tab:match_types_stats}), such as CM-MySQL$\Leftrightarrow$TPCH (80\% direct matches) and CO-Oracle$\Leftrightarrow$CM-MySQL (63\% direct matches). While Pairwise models still outperform ReMatch at \textit{@top-t}=1 in CM-MySQL$\to$ CO-Oracle (0.667 vs. 0.593), CO-Oracle$\to$CM-MySQL  (0.938 vs. 0.810), CM-MySQL$\to$TPCH (0.7 vs. 0.633), the learned Column-Table predictions provide marginal gains at higher \textit{@top-t} values than the similarity baseline. 

Considering the mean recall performance across all smaller target schema scenarios at \textit{@top-t}=3, Holistic(2) achieves 0.935 recall and consistently outperforms ReMatch and Pairwise models up to $+6\%$ as \textit{@top-t} increases. In summary, for this set of matching scenarios, we reduce the search space by $\approx 60\%$ (\textit{@top-t}=3) while maintaining $\geq 90\%$ ground truth matches.

\underline{Mean(Sakila$=$target)}: Secondly, we report the mean recall of all scenarios that predict Sakila's 16 target tables. The Column-Table prediction task is more complex, evidenced by the ReMatch baseline dropping to a starting recall of 0.446. 

The shared latent space enforcement in Pairwise and Holistic model types consistently outperform ReMatch and require lower \textit{top-t} cardinality for full recall. However, in contrast to the simpler scenarios, models trained without view augmentation perform better. The Pairwise(1) model achieves $+24\%$ higher recall at \textit{@top-t}=1 (0.551) than ReMatch. Interestingly, retrieval augmented columns at $d_{\text{max}}=2$ consistently degrade performance (e.g., Pairwise(2) drops to 0.482). We attribute this to Sakila's higher number of base tables introducing training complexity.

Furthermore, Pairwise outperforms Holistic for Sakila targets. Note that Sakila represents a sub-domain (Movie Rental) compared to the generic retail focus of Oracle, TPCH, and MySQL, in essence resulting in a more difficult matching task, therefore, the Holistic model suffers from negative transfer. For example, Sakila's columns in the \verb|FILM| and \verb|RENTAL| tables represent a minority class compared to analogous columns in \verb|PRODUCT| and \verb|ORDER| tables. Consequently, for analogous matches, pairwise Column-Table Learning without view-augmented columns tends to be most effective.

For the CO-Oracle$\to$Sakila scenario, all methods struggle to match Oracle's \verb|DELIVERY_ADDRESS| from table \verb|SHIPMENT| to Sakila's \verb|CITY| and \verb|COUNTRY| tables. In Sakila, city and country names are located multiple joins away from the customer concept (\verb|CUSTOMER| $\to$ \verb|ADDRESS| $\to$ \verb|CITY| $\to$ \verb|COUNTRY|). Since we capped the number of traversed tables at $d_{\text{max}}$=$2$, Sakila's model successfully contextualizes \verb|DELIVERY_ADDRESS| with the scope of \verb|CUSTOMER| $\Join$ \verb|ADDRESS| but is prevented from capturing the more distant \verb|CITY| and \verb|COUNTRY| context.

In summary, for the more complex scenarios for predicting Sakila's 16 target tables, we reduce the search space by $\approx 80\%$ (\textit{@top-t}=3) while maintaining $\geq 80\%$ ground truth matches.

\begin{table*}[h]
\centering
\caption{Recall Performance (Mean $\pm$ Std) by Schema Model Type for Column-Table Prediction (@\textit{top-t}).}
\label{tab:recall_performance}

\setlength{\tabcolsep}{2pt}
\resizebox{\textwidth}{!}{%
\begin{tabular}{clcccccccccccccccc} 
\toprule 
\multirow{2}{*}{\textbf{Source$\rightarrow$Target}} & \multirow{2}{*}{\specialcell{\textbf{Model Type}\\ $(d_\text{max})$}} &
\multicolumn{15}{c}{\textbf{\textit{@top-t}}}\\
& & \textbf{1} & \textbf{2} & \textbf{3} & \textbf{4} & \textbf{5} & \textbf{6} & \textbf{7} & \textbf{8} & \textbf{9} & \textbf{10} & \textbf{11} & \textbf{12} & \textbf{13} & \textbf{14} & \textbf{15} & \textbf{16} \\ 
\midrule
\midrule

\multirow{5}{*}{CM-MySQL $\rightarrow$ CO-Oracle} 
 & ReMatch & 0.593 & \textbf{0.963} & \textbf{1} & \textbf{1} & \textbf{1} & \textbf{1} & \textbf{1} & - & - & - & - & - & - & - & - & - \\
 & Pairwise(1) & \textbf{0.667} $\pm$ 0.0 & 0.7 $\pm$ 0.027 & \textit{0.963} $\pm$ 0.0 & \textbf{1} & \textbf{1} & \textbf{1} & \textbf{1} & - & - & - & - & - & - & - & - & - \\
 & Pairwise(2) & 0.515 $\pm$ 0.012 & 0.648 $\pm$ 0.059 & 0.807 $\pm$ 0.023 & \textbf{1} & \textbf{1} & \textbf{1} & \textbf{1} & - & - & - & - & - & - & - & - & - \\
 & Holistic(1) & \underline{0.663} $\pm$ 0.012 & \textit{0.707} $\pm$ 0.012 & \underline{0.989} $\pm$ 0.018 & \textbf{1} & \textbf{1} & \textbf{1} & \textbf{1} & - & - & - & - & - & - & - & - & - \\
 & Holistic(2) & \textit{0.626} $\pm$ 0.027 & \underline{0.733} $\pm$ 0.023 & 0.819 $\pm$ 0.054 & \textbf{1} & \textbf{1} & \textbf{1} & \textbf{1} & - & - & - & - & - & - & - & - & - \\
\midrule

\multirow{5}{*}{TPCH $\rightarrow$ CO-Oracle} 
 & ReMatch & 0.440 & 0.680 & 0.800 & 0.880 & 0.880 & 0.960 & \textbf{1} & - & - & - & - & - & - & - & - & - \\
 & Pairwise(1) & \underline{0.648} $\pm$ 0.017 & \underline{0.856} $\pm$ 0.034 & 0.912 $\pm$ 0.017 & 0.956 $\pm$ 0.013 & \underline{0.972} $\pm$ 0.019 & \textbf{1} & \textbf{1} & - & - & - & - & - & - & - & - & - \\
 & Pairwise(2) & \textit{0.596} $\pm$ 0.03 & 0.688 $\pm$ 0.017 & \underline{0.956} $\pm$ 0.013 & \textbf{1} & \textbf{1} & \textbf{1} & \textbf{1} & - & - & - & - & - & - & - & - & - \\
 & Holistic(1) & 0.516 $\pm$ 0.044 & \textit{0.792} $\pm$ 0.025 & \textbf{0.96} $\pm$ 0.0 & \textit{0.96} $\pm$ 0.0 & \textit{0.96} $\pm$ 0.0 & \textbf{1} & \textbf{1} & - & - & - & - & - & - & - & - & - \\
 & Holistic(2) & \textbf{0.796} $\pm$ 0.035 & \textbf{0.884} $\pm$ 0.013 & \textit{0.948} $\pm$ 0.019 & \underline{0.992} $\pm$ 0.017 & \textbf{1} & \textbf{1} & \textbf{1} & - & - & - & - & - & - & - & - & - \\
\midrule

\multirow{5}{*}{Sakila $\rightarrow$ CO-Oracle} 
 & ReMatch & 0.607 & 0.714 & 0.893 & \textit{0.964} & \underline{0.964} & \textbf{1} & \textbf{1} & - & - & - & - & - & - & - & - & - \\
 & Pairwise(1) & \underline{0.7} $\pm$ 0.018 & \textit{0.75} $\pm$ 0.029 & \textbf{0.993} $\pm$ 0.015 & \textbf{0.996} $\pm$ 0.011 & \textbf{1} & \textbf{1} & \textbf{1} & - & - & - & - & - & - & - & - & - \\
 & Pairwise(2) & \textbf{0.704} $\pm$ 0.017 & \underline{0.857} $\pm$ 0.051 & 0.921 $\pm$ 0.023 & \underline{0.971} $\pm$ 0.028 & \textbf{1} & \textbf{1} & \textbf{1} & - & - & - & - & - & - & - & - & - \\
 & Holistic(1) & 0.679 $\pm$ 0.029 & 0.711 $\pm$ 0.026 & \textit{0.932} $\pm$ 0.02 & 0.946 $\pm$ 0.019 & \underline{0.964} $\pm$ 0.0 & \textbf{1} & \textbf{1} & - & - & - & - & - & - & - & - & - \\
 & Holistic(2) & \textit{0.682} $\pm$ 0.046 & \textbf{0.9} $\pm$ 0.037 & \underline{0.957} $\pm$ 0.015 & \textit{0.964} $\pm$ 0.0 & \underline{0.964} $\pm$ 0.0 & \textbf{1} & \textbf{1} & - & - & - & - & - & - & - & - & - \\
 
\midrule

\multirow{5}{*}{CO-Oracle $\rightarrow$ CM-MySQL} 
 & ReMatch & 0.810 & \textit{0.952} & 0.952 & \underline{0.952} & \textbf{1} & \textbf{1} & \textbf{1} & \textbf{1} & - & - & - & - & - & - & - & - \\
 & Pairwise(1) & \textbf{0.938} $\pm$ 0.032 & \textbf{0.962} $\pm$ 0.02 & \underline{0.995} $\pm$ 0.015 & \textbf{1} & \textbf{1} & \textbf{1} & \textbf{1} & \textbf{1} & - & - & - & - & - & - & - & - \\
 & Pairwise(2) & 0.729 $\pm$ 0.032 & \textbf{0.962} $\pm$ 0.03 & \textbf{1} & \textbf{1} & \textbf{1} & \textbf{1} & \textbf{1} & \textbf{1} & - & - & - & - & - & - & - & - \\
 & Holistic(1) & \underline{0.871} $\pm$ 0.032 & \underline{0.957} $\pm$ 0.015 & 0.967 $\pm$ 0.023 & \textbf{1} & \textbf{1} & \textbf{1} & \textbf{1} & \textbf{1} & - & - & - & - & - & - & - & - \\
 & Holistic(2) & \textit{0.824} $\pm$ 0.068 & 0.948 $\pm$ 0.015 & \textit{0.986} $\pm$ 0.023 & \textbf{1} & \textbf{1} & \textbf{1} & \textbf{1} & \textbf{1} & - & - & - & - & - & - & - & - \\
\midrule

\multirow{5}{*}{TPCH $\rightarrow$ CM-MySQL} 
 & ReMatch & \textbf{0.667} & \textbf{0.917} & \textbf{0.958} & \textbf{1} & \textbf{1} & \textbf{1} & \textbf{1} & \textbf{1} & - & - & - & - & - & - & - & - \\
 & Pairwise(1) & 0.488 $\pm$ 0.02 & 0.579 $\pm$ 0.054 & 0.762 $\pm$ 0.056 & \textit{0.9} $\pm$ 0.053 & \underline{0.996} $\pm$ 0.013 & \textbf{1} & \textbf{1} & \textbf{1} & - & - & - & - & - & - & - & - \\
 & Pairwise(2) & 0.488 $\pm$ 0.02 & \underline{0.717} $\pm$ 0.051 & \textit{0.9} $\pm$ 0.022 & \textbf{1} & \textbf{1} & \textbf{1} & \textbf{1} & \textbf{1} & - & - & - & - & - & - & - & - \\
 & Holistic(1) & \textit{0.492} $\pm$ 0.018 & 0.629 $\pm$ 0.013 & 0.892 $\pm$ 0.045 & \underline{0.992} $\pm$ 0.018 & \textbf{1} & \textbf{1} & \textbf{1} & \textbf{1} & - & - & - & - & - & - & - & - \\
 & Holistic(2) & \underline{0.496} $\pm$ 0.013 & \textit{0.688} $\pm$ 0.029 & \underline{0.917} $\pm$ 0.0 & \textbf{1} & \textbf{1} & \textbf{1} & \textbf{1} & \textbf{1} & - & - & - & - & - & - & - & - \\
\midrule

\multirow{5}{*}{Sakila $\rightarrow$ CM-MySQL} 
 & ReMatch & 0.564 & 0.692 & 0.872 & 0.923 & 0.974 & \textbf{1} & \textbf{1} & \textbf{1} & - & - & - & - & - & - & - & - \\
 & Pairwise(1) & \textbf{0.667} $\pm$ 0.021 & \underline{0.879} $\pm$ 0.012 & 0.918 $\pm$ 0.011 & 0.933 $\pm$ 0.018 & \textit{0.992} $\pm$ 0.012 & \textbf{1} & \textbf{1} & \textbf{1} & - & - & - & - & - & - & - & - \\
 & Pairwise(2) & \textit{0.587} $\pm$ 0.057 & \textit{0.777} $\pm$ 0.017 & \textit{0.949} $\pm$ 0.0 & \textit{0.949} $\pm$ 0.0 & \underline{0.997} $\pm$ 0.008 & \textbf{1} & \textbf{1} & \textbf{1} & - & - & - & - & - & - & - & - \\
 & Holistic(1) & \underline{0.644} $\pm$ 0.019 & \textbf{0.897} $\pm$ 0.017 & \underline{0.964} $\pm$ 0.013 & \textbf{0.979} $\pm$ 0.011 & 0.987 $\pm$ 0.014 & \textbf{1} & \textbf{1} & \textbf{1} & - & - & - & - & - & - & - & - \\
 & Holistic(2) & 0.446 $\pm$ 0.013 & 0.767 $\pm$ 0.008 & \textbf{0.972} $\pm$ 0.008 & \underline{0.974} $\pm$ 0.0 & \textbf{1} & \textbf{1} & \textbf{1} & \textbf{1} & - & - & - & - & - & - & - & - \\
\midrule

\multirow{5}{*}{CO-Oracle $\rightarrow$ TPCH} 
 & ReMatch & 0.520 & 0.680 & 0.800 & \textit{0.920} & \textit{0.920} & 0.960 & \textbf{1} & \textbf{1} & - & - & - & - & - & - & - & - \\
 & Pairwise(1) & \underline{0.656} $\pm$ 0.021 & 0.804 $\pm$ 0.023 & \underline{0.952} $\pm$ 0.017 & \underline{0.96} $\pm$ 0.0 & \underline{0.96} $\pm$ 0.0 & \underline{0.996} $\pm$ 0.013 & \textbf{1} & \textbf{1} & - & - & - & - & - & - & - & - \\
 & Pairwise(2) & \textit{0.648} $\pm$ 0.041 & \textit{0.852} $\pm$ 0.027 & \textit{0.948} $\pm$ 0.019 & \underline{0.96} $\pm$ 0.0 & \underline{0.96} $\pm$ 0.0 & \textit{0.984} $\pm$ 0.021 & \textbf{1} & \textbf{1} & - & - & - & - & - & - & - & - \\
 & Holistic(1) & 0.64 $\pm$ 0.027 & \underline{0.868} $\pm$ 0.027 & \textbf{0.96} $\pm$ 0.0 & \underline{0.96} $\pm$ 0.0 & \underline{0.96} $\pm$ 0.0 & \textbf{1} & \textbf{1} & \textbf{1} & - & - & - & - & - & - & - & - \\
 & Holistic(2) & \textbf{0.776} $\pm$ 0.028 & \textbf{0.908} $\pm$ 0.019 & \textbf{0.96} $\pm$ 0.0 & \textbf{1} & \textbf{1} & \textbf{1} & \textbf{1} & \textbf{1} & - & - & - & - & - & - & - & - \\
\midrule

\multirow{5}{*}{CM-MySQL $\rightarrow$ TPCH} 
 & ReMatch & 0.633 & \textbf{0.900} & \underline{0.967} & \textit{0.967} & \textbf{1} & \textbf{1} & \textbf{1} & \textbf{1} & - & - & - & - & - & - & - & - \\
 & Pairwise(1) & \textit{0.64} $\pm$ 0.026 & 0.703 $\pm$ 0.011 & 0.857 $\pm$ 0.022 & 0.913 $\pm$ 0.032 & \textit{0.963} $\pm$ 0.033 & \underline{0.987} $\pm$ 0.017 & \textbf{1} & \textbf{1} & - & - & - & - & - & - & - & - \\
 & Pairwise(2) & \textbf{0.7} $\pm$ 0.022 & \textit{0.847} $\pm$ 0.017 & 0.9 $\pm$ 0.0 & 0.937 $\pm$ 0.011 & 0.96 $\pm$ 0.014 & \textit{0.977} $\pm$ 0.016 & \underline{0.993} $\pm$ 0.014 & \textbf{1} & - & - & - & - & - & - & - & - \\
 & Holistic(1) & 0.633 $\pm$ 0.0 & 0.773 $\pm$ 0.034 & \textit{0.91} $\pm$ 0.016 & \underline{0.97} $\pm$ 0.011 & \underline{0.983} $\pm$ 0.018 & \textbf{1} & \textbf{1} & \textbf{1} & - & - & - & - & - & - & - & - \\
 & Holistic(2) & \underline{0.68} $\pm$ 0.017 & \underline{0.89} $\pm$ 0.016 & \textbf{0.993} $\pm$ 0.014 & \textbf{1} & \textbf{1} & \textbf{1} & \textbf{1} & \textbf{1} & - & - & - & - & - & - & - & - \\
\midrule

\multirow{5}{*}{Sakila $\rightarrow$ TPCH} 
 & ReMatch & 0.324 & 0.588 & 0.735 & 0.882 & \textit{0.971} & \textbf{1} & \textbf{1} & \textbf{1} & - & - & - & - & - & - & - & - \\
 & Pairwise(1) & \underline{0.418} $\pm$ 0.023 & \underline{0.703} $\pm$ 0.058 & \textbf{0.894} $\pm$ 0.015 & \textit{0.979} $\pm$ 0.024 & \textbf{1} & \textbf{1} & \textbf{1} & \textbf{1} & - & - & - & - & - & - & - & - \\
 & Pairwise(2) & \textit{0.35} $\pm$ 0.022 & \textbf{0.721} $\pm$ 0.037 & \underline{0.876} $\pm$ 0.012 & \textbf{1} & \textbf{1} & \textbf{1} & \textbf{1} & \textbf{1} & - & - & - & - & - & - & - & - \\
 & Holistic(1) & \textbf{0.426} $\pm$ 0.029 & \textit{0.615} $\pm$ 0.038 & 0.765 $\pm$ 0.031 & 0.976 $\pm$ 0.012 & \underline{0.979} $\pm$ 0.014 & \textbf{1} & \textbf{1} & \textbf{1} & - & - & - & - & - & - & - & - \\
 & Holistic(2) & 0.329 $\pm$ 0.012 & 0.571 $\pm$ 0.021 & \textit{0.868} $\pm$ 0.021 & \underline{0.988} $\pm$ 0.015 & \textbf{1} & \textbf{1} & \textbf{1} & \textbf{1} & - & - & - & - & - & - & - & - \\
 
\midrule
\midrule

\multirow{5}{*}{\underline{Mean(Sakila$\neq$target)}} 
 & ReMatch & 0.573 & \underline{0.787} & 0.886 & 0.943 & 0.968 & 0.991 & \textbf{1} & \textbf{1} & - & - & - & - & - & - & - & - \\
 & Single(2) & 0.454 & 0.685 & 0.852 & 0.930 & 0.954 & 0.975 & \textit{0.995} & \textbf{1} & - & - & - & - & - & - & - & - \\
 & Pairwise(1) & \textbf{0.647} & 0.771 & \textit{0.916} & 0.960 & \textit{0.987} & \underline{0.998} & \textbf{1} & \textbf{1} & - & - & - & - & - & - & - & - \\
 & Pairwise(2) & 0.591 & \textit{0.785} & 0.918 & \underline{0.980} & \underline{0.991} & \textit{0.996} & \underline{0.999} & \textbf{1} & - & - & - & - & - & - & - & - \\
 & Holistic(1) & \textit{0.618} & 0.772 & \underline{0.926} & \textit{0.976} & 0.982 & \textbf{1} & \textbf{1} & \textbf{1} & - & - & - & - & - & - & - & - \\
 & Holistic(2) & \underline{0.628} & \textbf{0.81} & \textbf{0.935} & \textbf{0.991} & \textbf{0.996} & \textbf{1} & \textbf{1} & \textbf{1} & - & - & - & - & - & - & - & - \\

\midrule
\midrule

\multirow{5}{*}{CO-Oracle $\rightarrow$ Sakila} 
 & ReMatch & 0.429 & 0.619 & 0.619 & 0.619 & 0.762 & 0.810 & 0.857 & 0.857 & 0.857 & 0.905 & 0.952 & 0.952 & \underline{0.952} & \textbf{1} & \textbf{1} & \textbf{1} \\
 & Pairwise(1) & \textbf{0.624} $\pm$ 0.027 & \textit{0.7} $\pm$ 0.045 & \textbf{0.838} $\pm$ 0.033 & \textbf{0.857} $\pm$ 0.0 & \textbf{0.948} $\pm$ 0.027 & \textbf{0.957} $\pm$ 0.027 & \textbf{0.967} $\pm$ 0.023 & \textbf{0.986} $\pm$ 0.023 & \textbf{0.995} $\pm$ 0.015 & \textbf{0.995} $\pm$ 0.015 & \textbf{1} & \textbf{1} & \textbf{1} & \textbf{1} & \textbf{1} & \textbf{1} \\
 & Pairwise(2) & \textit{0.557} $\pm$ 0.023 & 0.605 $\pm$ 0.023 & \underline{0.786} $\pm$ 0.034 & \underline{0.81} $\pm$ 0.0 & \textit{0.838} $\pm$ 0.025 & \textit{0.867} $\pm$ 0.02 & \textit{0.91} $\pm$ 0.042 & \textit{0.938} $\pm$ 0.023 & \textit{0.948} $\pm$ 0.015 & \textit{0.957} $\pm$ 0.015 & \textit{0.99} $\pm$ 0.02 & \textbf{1} & \textbf{1} & \textbf{1} & \textbf{1} & \textbf{1} \\
 & Holistic(1) & \underline{0.581} $\pm$ 0.038 & \textbf{0.714} $\pm$ 0.022 & \textit{0.757} $\pm$ 0.035 & \textit{0.805} $\pm$ 0.015 & \underline{0.862} $\pm$ 0.042 & \underline{0.929} $\pm$ 0.034 & \underline{0.938} $\pm$ 0.023 & \underline{0.952} $\pm$ 0.0 & \underline{0.952} $\pm$ 0.0 & \underline{0.971} $\pm$ 0.025 & \underline{0.995} $\pm$ 0.015 & \underline{0.995} $\pm$ 0.015 & \textbf{1} & \textbf{1} & \textbf{1} & \textbf{1} \\
 & Holistic(2) & 0.524 $\pm$ 0.0 & \underline{0.71} $\pm$ 0.035 & 0.729 $\pm$ 0.039 & 0.767 $\pm$ 0.027 & 0.795 $\pm$ 0.023 & 0.814 $\pm$ 0.015 & 0.843 $\pm$ 0.032 & 0.876 $\pm$ 0.033 & 0.905 $\pm$ 0.0 & 0.91 $\pm$ 0.015 & 0.933 $\pm$ 0.033 & \textit{0.976} $\pm$ 0.04 & \textbf{1} & \textbf{1} & \textbf{1} & \textbf{1} \\
\midrule

\multirow{5}{*}{CM-MySQL $\rightarrow$ Sakila} 
 & ReMatch & \textit{0.538} & 0.718 & 0.769 & 0.795 & 0.821 & 0.821 & 0.897 & \textit{0.923} & \underline{0.923} & \underline{0.923} & \underline{0.923} & \underline{0.949} & \underline{0.949} & \textbf{1} & \textbf{1} & \textbf{1} \\
 & Pairwise(1) & \textbf{0.633} $\pm$ 0.057 & \textbf{0.874} $\pm$ 0.028 & \underline{0.897} $\pm$ 0.0 & \textit{0.931} $\pm$ 0.012 & \textit{0.946} $\pm$ 0.015 & \underline{0.972} $\pm$ 0.008 & \textit{0.985} $\pm$ 0.013 & \underline{0.997} $\pm$ 0.008 & \textbf{1} & \textbf{1} & \textbf{1} & \textbf{1} & \textbf{1} & \textbf{1} & \textbf{1} & \textbf{1} \\
 & Pairwise(2) & 0.533 $\pm$ 0.029 & \textit{0.828} $\pm$ 0.024 & \textit{0.895} $\pm$ 0.008 & \textbf{0.949} $\pm$ 0.0 & \underline{0.949} $\pm$ 0.0 & 0.949 $\pm$ 0.0 & 0.977 $\pm$ 0.015 & \underline{0.997} $\pm$ 0.008 & \textbf{1} & \textbf{1} & \textbf{1} & \textbf{1} & \textbf{1} & \textbf{1} & \textbf{1} & \textbf{1} \\
 & Holistic(1) & \underline{0.597} $\pm$ 0.047 & \underline{0.846} $\pm$ 0.024 & 0.892 $\pm$ 0.016 & 0.926 $\pm$ 0.008 & \textbf{0.951} $\pm$ 0.019 & \textbf{0.974} $\pm$ 0.012 & \underline{0.992} $\pm$ 0.012 & \underline{0.997} $\pm$ 0.008 & \textbf{1} & \textbf{1} & \textbf{1} & \textbf{1} & \textbf{1} & \textbf{1} & \textbf{1} & \textbf{1} \\
 & Holistic(2) & 0.423 $\pm$ 0.03 & \textit{0.828} $\pm$ 0.042 & \textbf{0.9} $\pm$ 0.008 & \underline{0.946} $\pm$ 0.008 & \textbf{0.951} $\pm$ 0.008 & \textit{0.969} $\pm$ 0.011 & \textbf{1} & \textbf{1} & \textbf{1} & \textbf{1} & \textbf{1} & \textbf{1} & \textbf{1} & \textbf{1} & \textbf{1} & \textbf{1} \\
\midrule

\multirow{5}{*}{TPCH $\rightarrow$ Sakila} 
 
 & ReMatch & 0.370 & \textit{0.593} & 0.667 & 0.741 & 0.741 & 0.741 & 0.741 & \textit{0.778} & \textit{0.852} & \textit{0.889} & \underline{0.889} & \underline{0.926} & \underline{0.926} & \textbf{1} & \textbf{1} & \textbf{1} \\
 & Pairwise(1) & \textbf{0.396} $\pm$ 0.039 & \textbf{0.678} $\pm$ 0.068 & \textbf{0.885} $\pm$ 0.027 & \textbf{0.941} $\pm$ 0.036 & \underline{0.97} $\pm$ 0.016 & 0.978 $\pm$ 0.019 & \textit{0.993} $\pm$ 0.016 & \underline{0.996} $\pm$ 0.012 & \underline{0.996} $\pm$ 0.012 & \underline{0.996} $\pm$ 0.012 & \textbf{1} & \textbf{1} & \textbf{1} & \textbf{1} & \textbf{1} & \textbf{1} \\
 & Pairwise(2) & 0.356 $\pm$ 0.05 & 0.578 $\pm$ 0.036 & \underline{0.837} $\pm$ 0.05 & \textbf{0.941} $\pm$ 0.031 & \textit{0.967} $\pm$ 0.027 & \textit{0.985} $\pm$ 0.019 & \underline{0.996} $\pm$ 0.012 & \textbf{1} & \textbf{1} & \textbf{1} & \textbf{1} & \textbf{1} & \textbf{1} & \textbf{1} & \textbf{1} & \textbf{1} \\
 & Holistic(1) & \underline{0.393} $\pm$ 0.047 & \textbf{0.678} $\pm$ 0.05 & 0.807 $\pm$ 0.057 & \textit{0.893} $\pm$ 0.027 & \textit{0.967} $\pm$ 0.021 & \underline{0.996} $\pm$ 0.012 & \textbf{1} & \textbf{1} & \textbf{1} & \textbf{1} & \textbf{1} & \textbf{1} & \textbf{1} & \textbf{1} & \textbf{1} & \textbf{1} \\
 & Holistic(2) & \textit{0.378} $\pm$ 0.052 & \underline{0.674} $\pm$ 0.046 & \textit{0.822} $\pm$ 0.034 & \underline{0.93} $\pm$ 0.037 & \textbf{0.978} $\pm$ 0.026 & \textbf{1} & \textbf{1} & \textbf{1} & \textbf{1} & \textbf{1} & \textbf{1} & \textbf{1} & \textbf{1} & \textbf{1} & \textbf{1} & \textbf{1} \\
\midrule
\midrule

\multirow{5}{*}{\underline{Mean(Sakila$=$target)}} 
 & ReMatch & 0.446 & 0.643 & 0.685 & 0.718 & 0.774 & 0.790 & 0.832 & 0.853 & 0.877 & 0.906 & 0.921 & 0.942 & 0.942 & \underline{0.942} & \textbf{1} & \textbf{1} \\
 & Single(2) & 0.291 & 0.509 & 0.657 & 0.783 & 0.889 & \textit{0.943} & \textit{0.97} & \textit{0.979} & \underline{0.989} & \textbf{0.998} & \textbf{1} & \textbf{1} & \textbf{1} & \textbf{1} & \textbf{1} & \textbf{1} \\
 & Pairwise(1) & \textbf{0.551} & \textbf{0.751} & \textbf{0.874} & \textbf{0.91} & \textbf{0.955} & \textbf{0.969} & \textbf{0.981} & \textbf{0.993} & \textbf{0.997} & \underline{0.997} & \textbf{1} & \textbf{1} & \textbf{1} & \textbf{1} & \textbf{1} & \textbf{1} \\
 & Pairwise(2) & \textit{0.482} & 0.67 & \underline{0.839} & \underline{0.9} & \textit{0.918} & 0.934 & 0.961 & \textit{0.979} & 0.983 & 0.986 & \textit{0.997} & \textbf{1} & \textbf{1} & \textbf{1} & \textbf{1} & \textbf{1} \\
 & Holistic(1) & \underline{0.524} & \underline{0.746} & \textit{0.819} & 0.874 & \underline{0.927} & \underline{0.966} & \underline{0.977} & \underline{0.983} & \textit{0.984} & \textit{0.990} & \underline{0.998} & \underline{0.998} & \textbf{1} & \textbf{1} & \textbf{1} & \textbf{1} \\
 & Holistic(2) & 0.442 & \textit{0.737} & 0.817 & \textit{0.881} & 0.908 & 0.928 & 0.948 & 0.959 & 0.968 & 0.970 & 0.978 & \textit{0.992} & \textbf{1} & \textbf{1} & \textbf{1} & \textbf{1} \\

\midrule

\multirow{5}{*}{\underline{Mean(all)}} 
 & ReMatch & 0.541 & 0.751 & 0.836 & 0.887 & 0.919 & 0.941 & 0.958 & 0.951 & 0.877 & 0.906 & 0.921 & 0.942 & 0.942 & \underline{0.942} & \textbf{1} & \textbf{1} \\
 & Single(2) & 0.413 & 0.641 & 0.804 & 0.893 & 0.938 & 0.967 & 0.989 & \textit{0.993} & \underline{0.989} & \textbf{0.998} & \textbf{1} & \textbf{1} & \textbf{1} & \textbf{1} & \textbf{1} & \textbf{1} \\
 & Pairwise(1) & \textbf{0.623} & \underline{0.766} & \textbf{0.906} & 0.947 & \textbf{0.979} & \underline{0.991} & \textbf{0.995} & \textbf{0.998} & \textbf{0.997} & \underline{0.997} & \textbf{1} & \textbf{1} & \textbf{1} & \textbf{1} & \textbf{1} & \textbf{1} \\
 & Pairwise(2) & 0.563 & \textit{0.757} & \textit{0.898} & \underline{0.96} & \textit{0.973} & 0.980 & \textit{0.990} & \textit{0.993} & 0.983 & 0.986 & \textit{0.997} & \textbf{1} & \textbf{1} & \textbf{1} & \textbf{1} & \textbf{1} \\
 & Holistic(1) & \underline{0.595} & \underline{0.766} & \underline{0.900} & \textit{0.951} & 0.968 & \textbf{0.992} & \underline{0.994} & \underline{0.994} & \textit{0.984} & \textit{0.990} & \underline{0.998} & \underline{0.998} & \textbf{1} & \textbf{1} & \textbf{1} & \textbf{1} \\
 & Holistic(2) & \textit{0.582} & \textbf{0.792} & \textbf{0.906} & \textbf{0.963} & \underline{0.974} & \textit{0.982} & 0.987 & 0.986 & 0.968 & 0.970 & 0.978 & \textit{0.992} & \textbf{1} & \textbf{1} & \textbf{1} & \textbf{1} \\

\bottomrule
\multicolumn{18}{l}{All RACT models use Retrieval Augmented Columns (ref. Algorithm \ref{alg:view_augmentation} $\oplus$) with $dir = \text{directed}$, uniform $d_\text{max}$ value in parenthesis, and $\lambda = 0.5$. ReMatch uses Cosine similarity between column ($Ser_{\textit{schema}}$) and table ($Ser_{\textit{ReMatch}}^t$}) embeddings. \\
\multicolumn{18}{l}{The best recall per model type are formatted in \textbf{bold}, the second-best in \underline{underlined}, and the third-best in \textit{italic} font.}\\\ 

\end{tabular}%
}
\end{table*}

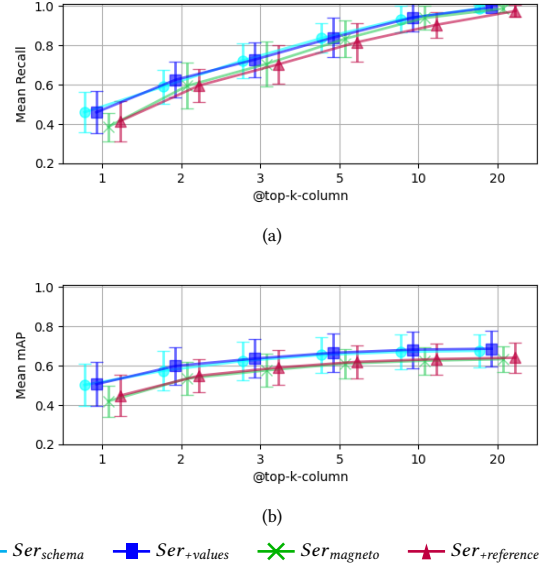
\begin{figure}[t]
  \centering  
  \subfigure[]{\includegraphics[width=.4\textwidth]{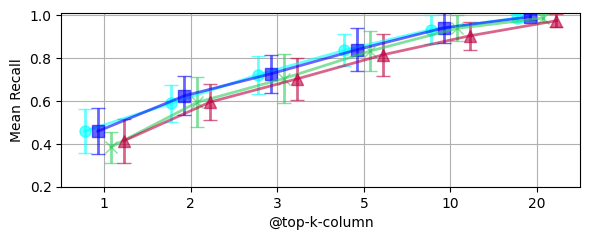}}
  \subfigure[]{\includegraphics[width=.4\textwidth]{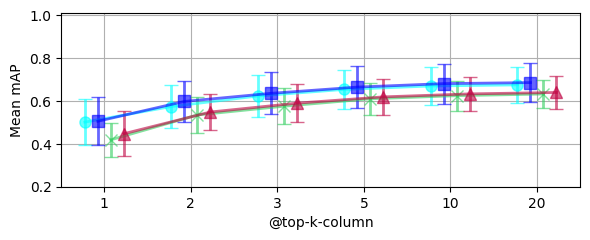}}
  
  \begin{subfigure}
    \centering
    \footnotesize
    \begin{tabular}{@{}*{9}{l}@{}}
      \tikz[baseline=-0.6ex]{
          \draw[cyan,solid,line width=1pt](0,0)--(2em,0); \fill[cyan] (1em,0) circle (3pt);} $Ser_{\textit{schema}}$ &
      
      \tikz[baseline=-0.6ex]{
          \draw[blue,solid,line width=1pt](0,0)--(2em,0); \fill[blue] (0.7em,-3pt) rectangle (1.3em,3pt);} $Ser_{\textit{+values}}$ &
          
      \tikz[baseline=-0.6ex]{
           \draw[green!70!black,solid,line width=1pt](0,0)--(2em,0); 
           \draw[green!70!black,line width=1pt] (1em-3pt,-3pt)--(1em+3pt,3pt) (1em-3pt,3pt)--(1em+3pt,-3pt);} $Ser_{\textit{magneto}}$ &
      
      \tikz[baseline=-0.6ex]{
          \draw[purple,solid,line width=1pt](0,0)--(2em,0); \fill[purple] (0.75em,-3pt) -- (1em,3pt) -- (1.25em,-3pt) -- cycle; }  $Ser_{\textit{+reference}}$ &
    \end{tabular}
  \end{subfigure}
  \caption{Impact of Serialization for Column Blocking (\textit{@top-k}) and Matching (\textit{Cosine}) measured in Recall (a) and mAP (b) Performance as Mean over all Scenarios.}
  \label{fig:serialization_ablation}
\end{figure}

\begin{figure*}
  \vspace{-2em}
  \centering  
  \subfigure[CM-MySQL $\rightarrow$ CO-Oracle]{\includegraphics[width=0.245\textwidth]{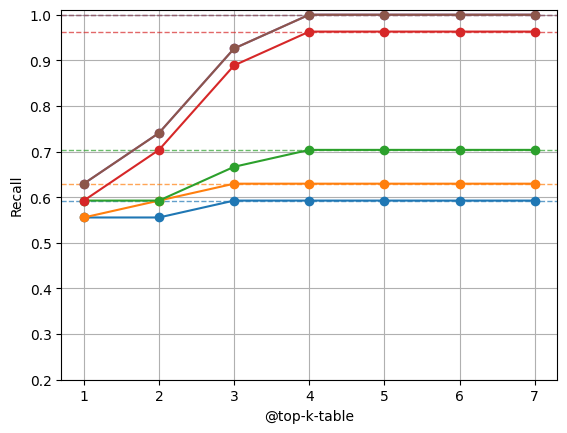}}
  \subfigure[CO-Oracle $\rightarrow$ CM-MySQL]{\includegraphics[width=0.245\textwidth]{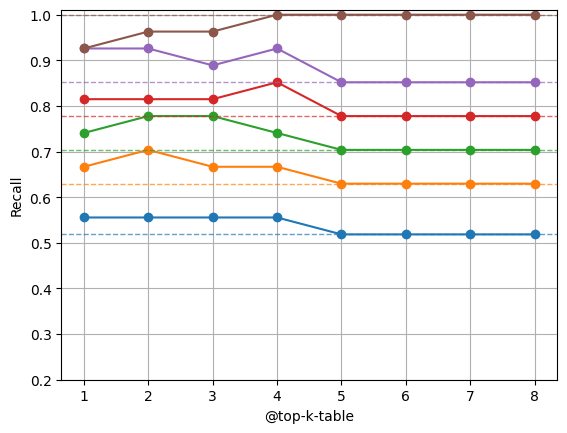}}
  \subfigure[CO-Oracle $\rightarrow$ TPCH]{\includegraphics[width=0.245\textwidth]{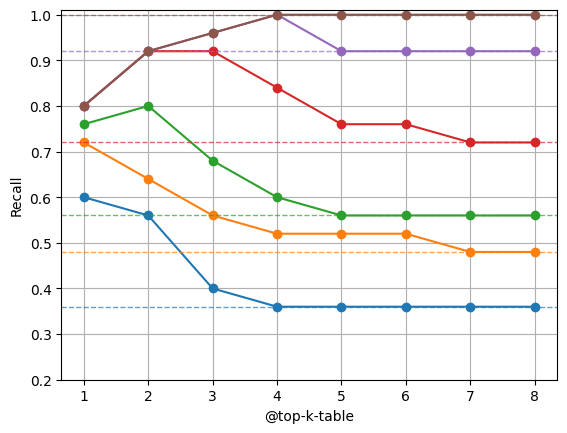}}
  \subfigure[CO-Oracle $\rightarrow$ Sakila]{\includegraphics[width=0.245\textwidth]{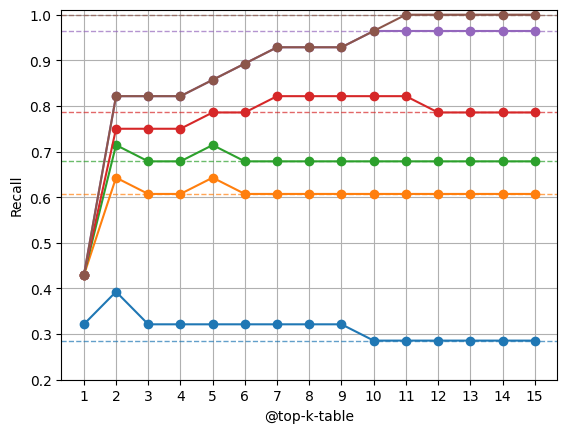}}
  \subfigure[TPCH $\rightarrow$ CO-Oracle]{\includegraphics[width=0.245\textwidth]{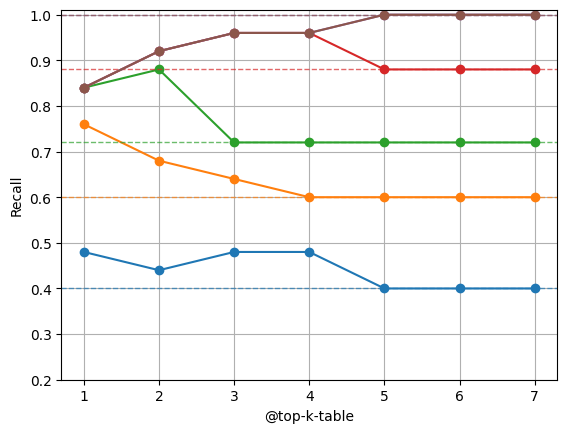}}
  \subfigure[TPCH $\rightarrow$ CM-MySQL]{\includegraphics[width=0.245\textwidth]{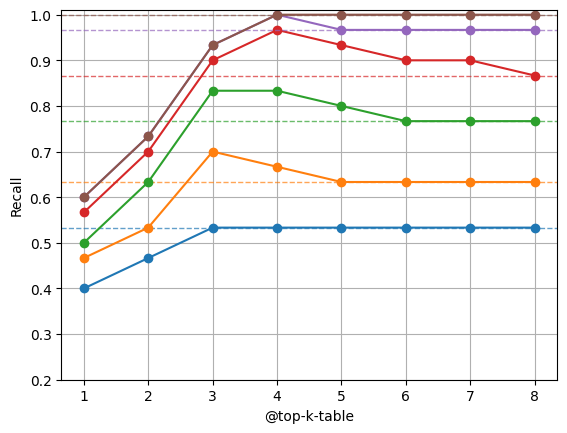}}
  \subfigure[CM-MySQL $\rightarrow$ TPCH]{\includegraphics[width=0.245\textwidth]{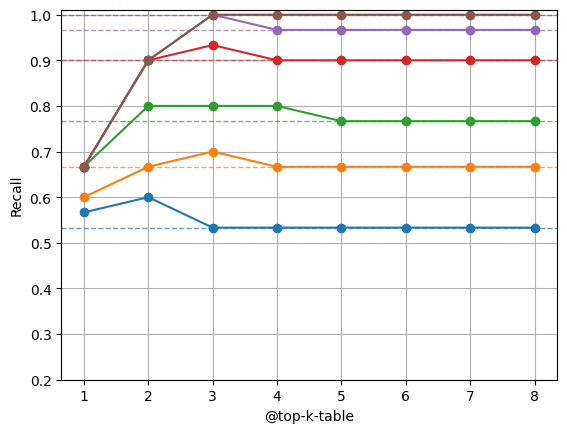}}
  \subfigure[CM-MySQL $\rightarrow$ Sakila]{\includegraphics[width=0.245\textwidth]{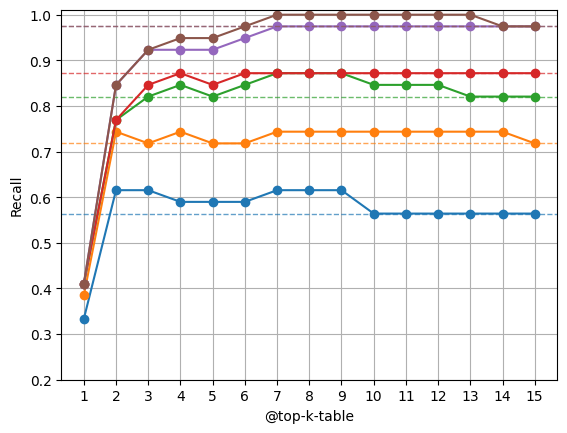}}
  \subfigure[Sakila $\rightarrow$ CO-Oracle]{\includegraphics[width=0.245\textwidth]{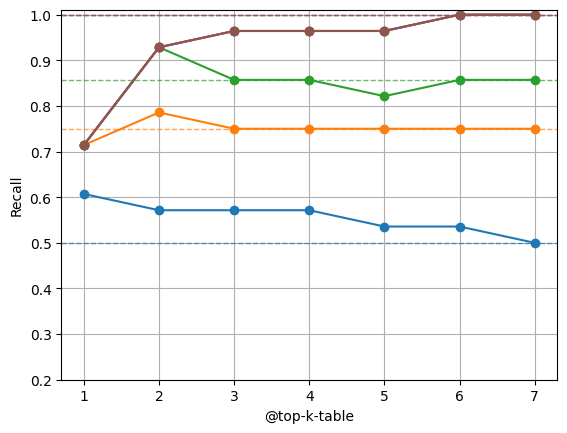}}
  \subfigure[Sakila $\rightarrow$ CM-MySQL]{\includegraphics[width=0.245\textwidth]{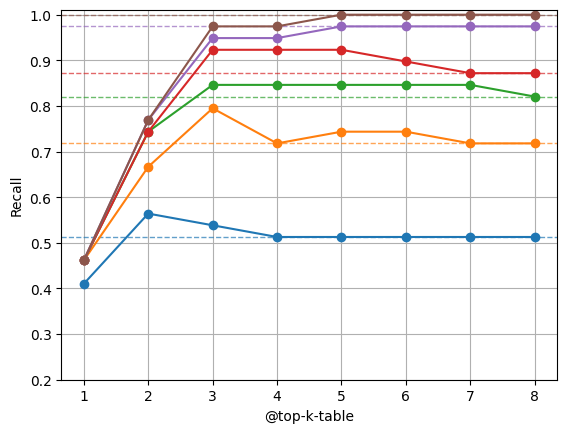}}
  \subfigure[Sakila $\rightarrow$ TPCH]{\includegraphics[width=0.245\textwidth]{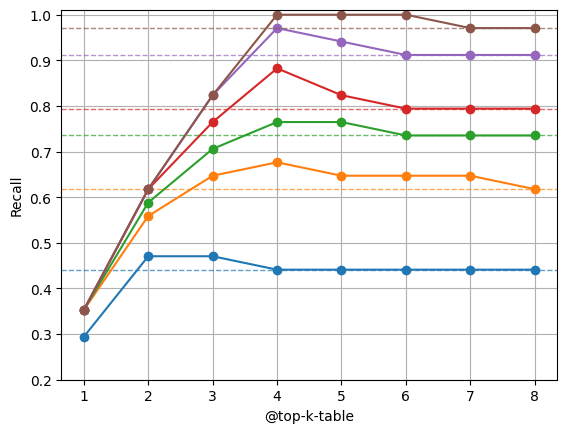}}
  \subfigure[TPCH $\rightarrow$ Sakila]{\includegraphics[width=0.245\textwidth]{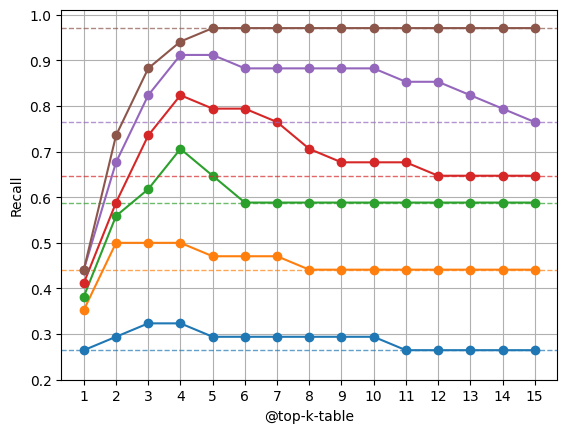}}
  \begin{subfigure}
    \centering
    \footnotesize 
    \begin{tabular}{@{}*{9}{l}@{}}
      \tikz[baseline=-0.6ex]{\draw[NavyBlue,solid,line width=1pt](0,0)--(2em,0);} @top-k=1&
      \tikz[baseline=-0.6ex]{\draw[orange,solid,line width=1pt](0,0)--(2em,0);} @top-k=2 &
      \tikz[baseline=-0.6ex]{\draw[OliveGreen,solid,line width=1pt](0,0)--(2em,0);} @top-k=3 &
      \tikz[baseline=-0.6ex]{\draw[OrangeRed,solid,line width=1pt](0,0)--(2em,0);} @top-k=5 &
      \tikz[baseline=-0.6ex]{\draw[RoyalPurple,solid,line width=1pt](0,0)--(2em,0);} @top-k=10 &
      \tikz[baseline=-0.6ex]{\draw[Brown,solid,line width=1pt](0,0)--(2em,0);} @top-k=20
    \end{tabular}
  \end{subfigure}
  
  \caption{Recall Performance (y-axis) at \textit{@top-t} Table Prediction (x-axis) with Holistic(2) and Blocking \textit{@top-k} Column (lines).}
  \label{fig:blocking_results}
\end{figure*}

\begin{figure*}
  \centering  
  \subfigure[CM-MySQL $\rightarrow$ CO-Oracle]{\includegraphics[width=0.245\textwidth]{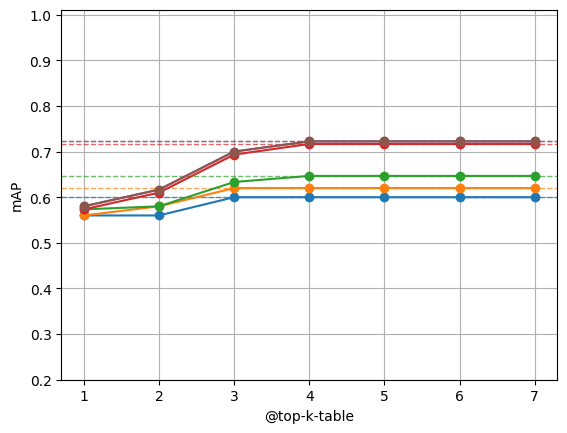}}
  \subfigure[CO-Oracle $\rightarrow$ CM-MySQL]{\includegraphics[width=0.245\textwidth]{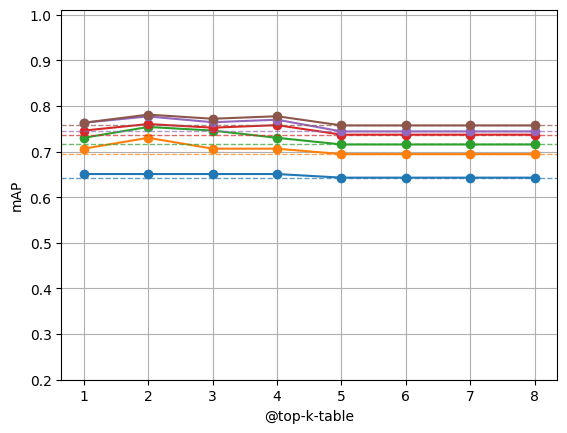}}
  \subfigure[CO-Oracle $\rightarrow$ TPCH]{\includegraphics[width=0.245\textwidth]{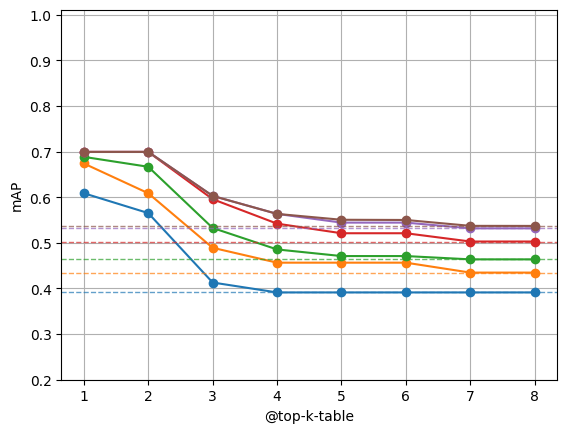}}
  \subfigure[CO-Oracle $\rightarrow$ Sakila]{\includegraphics[width=0.245\textwidth]{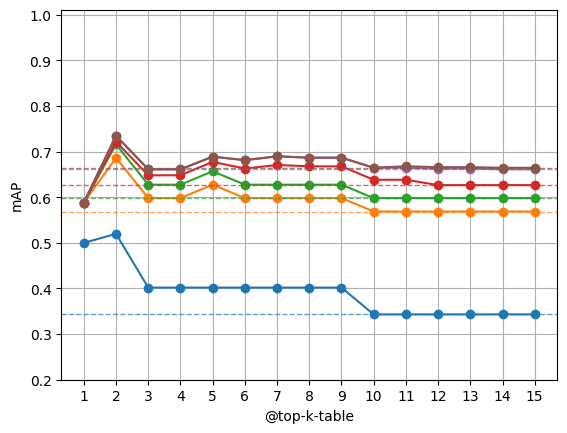}}
  \subfigure[TPCH $\rightarrow$ CO-Oracle]{\includegraphics[width=0.245\textwidth]{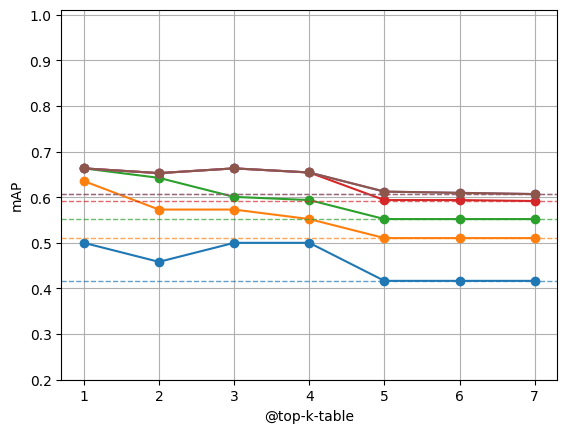}}
  \subfigure[TPCH $\rightarrow$ CM-MySQL]{\includegraphics[width=0.245\textwidth]{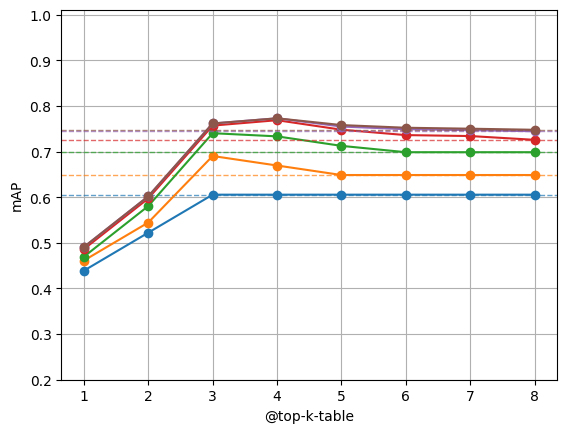}}
  \subfigure[CM-MySQL $\rightarrow$ TPCH]{\includegraphics[width=0.245\textwidth]{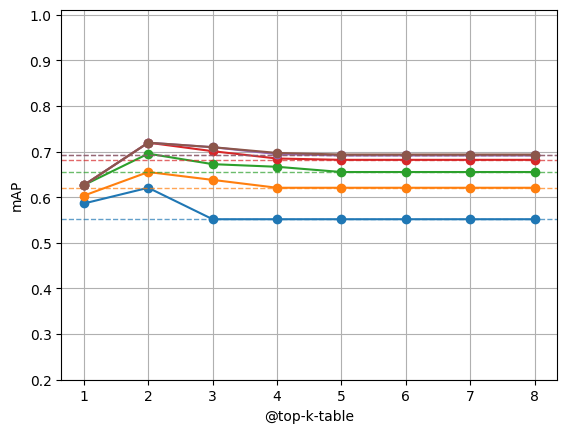}}
  \subfigure[CM-MySQL $\rightarrow$ Sakila]{\includegraphics[width=0.245\textwidth]{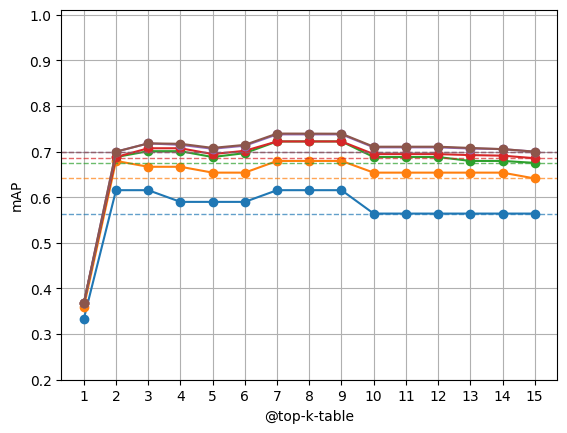}}
  \subfigure[Sakila $\rightarrow$ CO-Oracle]{\includegraphics[width=0.245\textwidth]{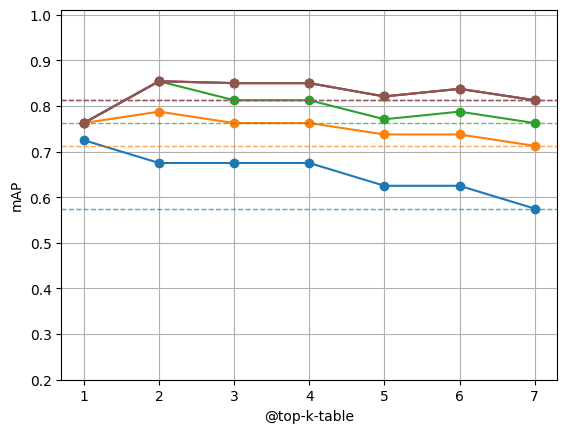}}
  \subfigure[Sakila $\rightarrow$ CM-MySQL]{\includegraphics[width=0.245\textwidth]{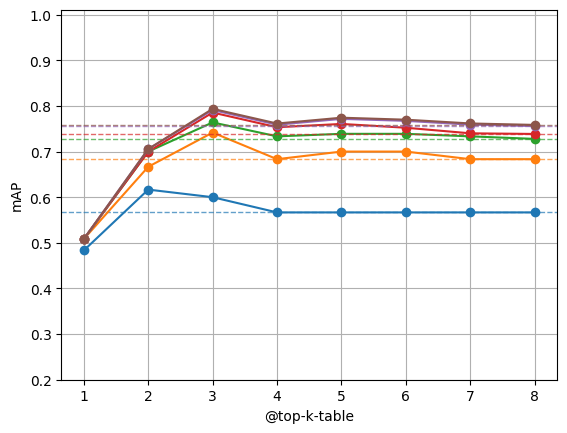}}
  \subfigure[Sakila $\rightarrow$ TPCH]{\includegraphics[width=0.245\textwidth]{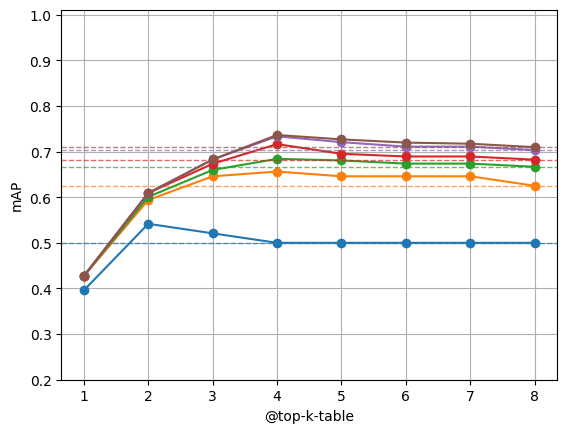}}
  \subfigure[TPCH $\rightarrow$ Sakila]{\includegraphics[width=0.245\textwidth]{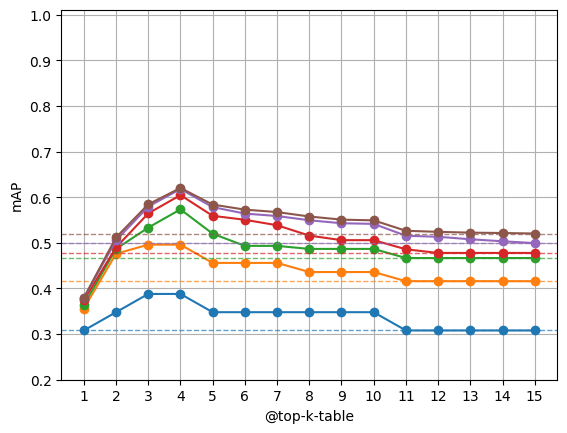}}

  \begin{subfigure}
    \centering
    \footnotesize 
    \begin{tabular}{@{}*{9}{l}@{}}
      \tikz[baseline=-0.6ex]{\draw[NavyBlue,solid,line width=1pt](0,0)--(2em,0);} @top-k=1&
      \tikz[baseline=-0.6ex]{\draw[orange,solid,line width=1pt](0,0)--(2em,0);} @top-k=2 &
      \tikz[baseline=-0.6ex]{\draw[OliveGreen,solid,line width=1pt](0,0)--(2em,0);} @top-k=3 &
      \tikz[baseline=-0.6ex]{\draw[OrangeRed,solid,line width=1pt](0,0)--(2em,0);} @top-k=5 &
      \tikz[baseline=-0.6ex]{\draw[RoyalPurple,solid,line width=1pt](0,0)--(2em,0);} @top-k=10 &
      \tikz[baseline=-0.6ex]{\draw[Brown,solid,line width=1pt](0,0)--(2em,0);} @top-k=20
    \end{tabular}
  \end{subfigure}
  
  \caption{mAP Performance (y-axis) at \textit{@top-k} Table Prediction (x-axis) with Holistic(2) and Matching \textit{@top-k} Column (lines).}
  \label{fig:map_results}
\end{figure*}

\underline{Mean(all)}: Lastly, we report the recall aggregates across all scenarios to identify a default strategy. At \textit{@top-t}=1, the Pairwise model without augmentation yields the best recall with 0.623. The Holistic(2) model emerges as the strongest overall approach, recovering from the false positive tables predicted at \textit{@top-t}=1. Specifically, it overtakes all other model types as well as ReMatch at \textit{@top-t}=2 (0.792), \textit{@top-t}=3 (0.906), and \textit{@top-t}=4 (0.963) until all model types converge as \textit{@top-t} increases. Overall, the similarity-based ReMatch baseline consistently underperforms compared to Pairwise and Holistic models across the \textit{top-t} table spectrum.

\textbf{Impact of Serialization for Schema Matching.}\\ $Ser_{\textit{+reference}}$ yielded higher validation accuracy and recall on Column-Table Learning. Intuitively, encoding schema context via referential information allows the model to learn the table context more effectively. However, the added referential context (via adding \textit{``Weak Table''} and \textit{``Strong Table''}) may not benefit fine-grained column matching. Therefore, we design a study to evaluate the impact of the four different serialization variants using column Blocking (\textit{top-k}) and Matching (Cosine similarity). 

In Figure \ref{fig:serialization_ablation}, we report the mean recall (y-axis in a) and mean mAP (y-axis in b) over all scenarios at \textit{top-k} columns (x-axis) with $Ser_{\textit{schema}}$ (dot-cyan), $Ser_{\textit{+values}}$ (square-blue), $Ser_{\textit{magneto}}$ (cross-green), and $Ser_{\textit{+reference}}$ (triangle-purple) as line plots. 

At any \textit{top-k} column cardinality, $Ser_{\textit{+values}}$ ranks the best closely followed by $Ser_{\textit{schema}}$ for both recall and mAP, confirming our table extension to \cite{liu_magneto_2025}. 
While $Ser_{\textit{schema}}$ and $Ser_{\textit{+values}}$ perform similarly, there is an evident gap ($0.05$ to $0.1$) to the mAP and recall performance of $Ser_{\textit{magneto}}$ and $Ser_{\textit{+reference}}$. Notably, $Ser_{\textit{magneto}}$ has higher recall deviations among the scenarios (error-bar) than others. Particularly at \textit{@top-k}=1, we attribute $Ser_{\textit{magneto}}$ weaker mAP to the missing table context needed to preciously identify matches. For $Ser_{\textit{+reference}}$, on the other hand, the added referential context overlays column semantics and leads to embedding noise instead of matching signal. Hence, referential context is crucial for Column-Table learning while harming column-column similarity. 

\textbf{Ablation Study for Schema Matching.} Based on our previous findings, we evaluate the impact of Column-Table Prediction on the full matching pipeline. First, we train a Holistic model for Column-Table Prediction with $d_{\text{max}}=2$ augmentation of $Ser_{\textit{+reference}}$ serialized column embeddings. Then, we apply classical Blocking and Matching using $Ser_{\textit{+values}}$ serialized column embeddings. In Figure \ref{fig:blocking_results}, we report the recall performance (y-axis) for each scenario. The x-axis represents Column-Table Prediction at \textit{top-t} candidates, while the colored lines \{1 (cyan), 2 (orange), 3 (green), 5 (red), 10 (purple), 20 (brown)\} represent Blocking with \textit{top-k} column cardinality. Finally, the dashed horizontal lines indicate the baseline performance (maximum \textit{@top-t} corresponds to unconstrained search) for each \textit{top-k} column level. Correspondingly, we report mAP performance in Figure \ref{fig:map_results}. As already observed by Bellahsene et al. \cite{bellahsene_evaluating_2011}, also each matching scenario in our dataset represents a uniquely challenging solution space. However, we observe the following patterns with our Column-Table Prediction that precedes classical similarity-based matching:

\underline{Matching Quality.} All scenarios (except a) can benefit from a \textit{top-t} table constrained search space yielding higher recall at several \textit{top-k} column cardinalities compared to the full search baseline. Particularly for scenarios (c) and (e), recall and mAP improve by up to $+70\%$ @top-k=1-10. For @top-k=20 blocking, recall performance becomes comparable at @top-t=4/5 candidates. The only exception is scenario (d), reaching on-par recall @top-t=10 that we relate to the previously discussed high normalization of \verb|ADDRESSES|, \verb|CITY|, and \verb|COUNTRY| tables. 

\underline{Contextual Alignment.} Generally, scenarios with larger shares of contextual matches (d and i), (h and j), and (k and l) struggle at initial \textit{top-t} table candidates in recall but recover at @top-t=3/4 tables while steadily surpassing baseline mAP performance. With RACT prediction, the contextual matches between Sakila and TPCH (i.e., \verb|FILM|$=$\verb|PART| and \verb|RENTAL|$=$\verb|ORDER|) are revealed by up to $+28\%$ recall for scenarios (k) and (l). Notably, scenarios (h) and (k) reach full recall at @top-k=20 column blocking only with a \textit{top-t} table constraint. This effect demonstrates that the similarity-based spaces can benefit from referential context for matching relational schemas.  

\underline{Setting \textit{top-t}.} We observe that \textit{@top-t=4/5} candidates perform on-par or better among most matching scenarios at any \textit{top-k} column blocking, aligning with our intrinsic method analysis (Table \ref{tab:recall_performance}). Generally, we advise lower \textit{top-t} table constraints for lower \textit{top-k} column blocking. On the other hand, larger \textit{top-k} column blocking also benefits from larger \textit{top-t} table constraints. 

\underline{Limitations.} First, Column-Table Learning and Prediction is a probabilistic approximation that does not guarantee retrieving accurate host tables at low \textit{top-t} table cardinalities. Secondly, our approach heavily relies on schema metadata and referential constraints (provided in the RACT dataset), which, in general, are obtainable but may sometimes be incomplete or unavailable. 

\underline{Explainability.} Looking beyond \textit{top-t} constraints as an optimization problem, including the table concepts via column-table candidate retrieval offers practical advantages for human-in-the-loop and LLM frameworks. Unlike opaque similarity scores between columns, RACT recommends semantically meaningful target table names, enabling practitioners (or agents) to intuitively verify and steer the search space in holistic multi-table matching scenarios.

\section{Conclusion}

In this paper, we discussed the problem of identifying matches between multiple tables in different contexts. Our RACT framework aims to solve this problem by generating directed schema graphs from relational schemas and learning the referential table context of a column using self-supervised neural network models. Subsequently, given a column in a schema, the learned RACT model predicts appropriate candidate tables of another schema. This approach significantly differs from matching methods that apply similarity-based techniques, since RACT models simultaneously learn probabilistic weights in a shared latent space among all candidate schemas. 
Evaluations show that our approach is more effective for scenarios containing contextual matches compared to similarity-based approaches for both table retrieval and column matching alone. At the same time, the number of models needed to train among multiple schemas reduces to a single holistically shared RACT model. In the future, we plan to extend RACT with column augmentation that is table-class balanced and weighted join paths in schema graphs.
%


%


\section{Artifacts}

All experiments were conducted in a Python Jupyter Notebook on an Intel i7-1265U CPU with 32GB memory. All relevant datasets for reproducing the experiments including the relational schemas, referential constraints, annotated ground truth matches, reported experimental results, and the executable Python Jupyter notebook \verb|RACT.ipynb| are publicly accessible without monitoring in the GitHub repository \url{https://github.com/leotraeg/RACT}. A thorough description of the datasets and their origin, as well as a quick-start description of the algorithmic implementation, including performance metrics, is provided in the \verb|README.md| file.

\bibliographystyle{ACM-Reference-Format}

\bibliography{literature.bib}

\appendix

\clearpage

\end{document}